\documentclass[sigconf, screen, nonacm]{acmart}
\usepackage{braket}
\usepackage{subcaption}
\usepackage{bm}

\AtBeginDocument{%
  }
    
\setcopyright{none} 
\newcommand{\acronym}{our framework}

\begin{document}

\title{Logical Compilation for Multi-Qubit Iceberg Patches}

\author{Cordell Mazzetti}
\orcid{0009-0002-2350-6007}
\email{cmm7848@eid.utexas.edu}
\affiliation{%
  \institution{The University of Texas at Austin}
  \city{Austin}
  \state{Texas}
  \country{}
}

\author{Sayam Sethi}
\orcid{0009-0005-3056-5285}
\email{sayams@utexas.edu}
\affiliation{%
  \institution{The University of Texas at Austin}
  \city{Austin}
  \state{Texas}
  \country{}
}

\author{Rich Rines}
\orcid{0009-0005-4073-1291}
\affiliation{%
  \institution{Infleqtion}
  \city{Chicago}
  \state{Illinois}
  \country{}
}

\author{Pranav Gokhale}
\orcid{0000-0003-1946-4537}
\affiliation{%
  \institution{Infleqtion}
  \city{Chicago}
  \state{Illinois}
  \country{}
}

\author{Jonathan Mark Baker}
\orcid{0000-0002-0775-8274}
\affiliation{%
  \institution{The University of Texas at Austin}
  \city{Austin}
  \state{Texas}
  \country{}
}

\begin{abstract}
Recent advancements in quantum computing have enabled practical use of quantum error detecting and correcting codes. However, current architectures and future proposals of quantum computer design suffer from limited qubit counts, necessitating the use of high-rate codes. Such codes, with their code parameters denoted as $[[n, k, d]]$, have more than $1$ logical qubit per code (i.e., $k > 1$). This leads to reduced error tolerance of the code, since $\lceil (d-1)/2\rceil$ errors on any of the $n$ physical qubits can affect the logical state of all $k$ logical qubits. Therefore, it becomes critical to optimally map the input qubits of a quantum circuit to these codes, in such a way that the circuit fidelity is maximized.
\par However, the problem of mapping program qubits to logical qubits for high-rate codes has not been studied in prior work. A brute force search to find the optimal mapping is super exponential (scaling as $O(n!)$, where $n$ is the number of input qubits), making exhaustive search infeasible past a small number of qubits. We propose a framework that addresses this problem on two fronts: (1) for any given mapping, it performs logical-to-physical compilation that translates input gates into efficiently encoded implementations utilizing Hadamard commutation and gate merging; and (2) it quickly searches the space of possible mappings through a merge-optimizing, noise-biased packing heuristic that identifies high-performing qubit assignments without exhaustive enumeration. To the best of our knowledge, \acronym{} is the first work to explore mapping and compilation for high-rate codes. Across 71 benchmark circuits, we reduce circuit depth by $34\%$, gate counts by up to $31\%$ and $17\%$ for one-qubit and two-qubit gates, and improve total variation distance by $1.75\times$, with logical selection rate improvements averaging $86\%$ relative to naive compilation.
\end{abstract}

\maketitle

\section{Introduction}
Quantum computing systems are rapidly improving with physical error rates dropping and qubit counts increasing\cite{manetsch2025tweezer}. Despite these improvements, important applications remain out of reach such as Shor's algorithm\cite{gidney_how_2021}. In order to obtain the error rates necessary to execute these programs reliably, one popular strategy is to employ some form of quantum error correction (QEC) which uses many physical qubits to encode individual logical states which can exponentially suppress error rates. Even still, to employ QEC codes can require huge numbers of physical qubits, far beyond what is currently (or even projected) to be available in the near to intermediate future. For example, popular codes like the surface code\cite{fowler2012surface, horsman2012surface, litinski_game_2018} require a quadratic increase in physical qubits to increase the distance by 1, which approximately quantifies how many physical errors can be detected and corrected by the code.

\begin{figure}
    \centering
    \includegraphics[width=1\linewidth]{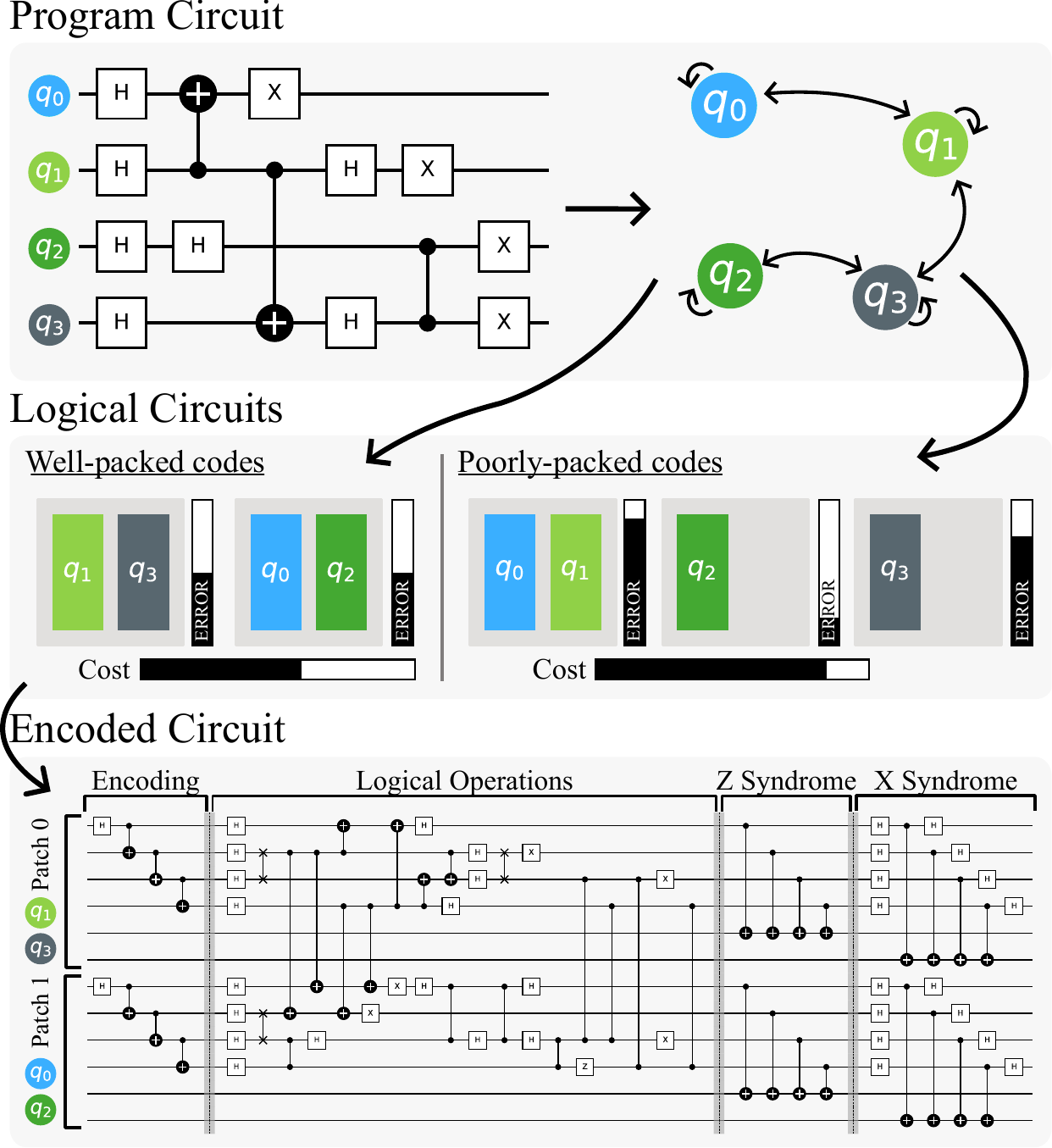}
    \caption{\textbf{Overview of the optimization problem.} Interactions between program qubits in the input circuit are evaluated to compile a permutation of logical qubits that minimizes the encoded circuit's resource overhead and maximizes output quality. A patch denotes a set of physical qubits that encode a set of logical qubits.}
    \label{fig:pipeline_flow}
\end{figure}

In recent years, industry has turned its attention from NISQ program execution towards the preparation for the fault tolerant quantum computing (FTQC) era\cite{Xuetal2024, Bravyietal2024, Ye2025, gidney_yoked_2023, Paetznicketal2024}. Ambitious roadmaps from nearly every company regardless of physical qubit implementation have laid out their plans for \textit{hardware} supported QEC with thousands of qubits \cite{Manetsch_Nomura_Bataille_Lv_Leung_Endres_2025}. In parallel, it remains critical to develop software-architectural support for these emerging systems which maximizes the quality of qubits available to us, for example by developing validation or compilation pipelines for both the future systems as well as intermediary steps. 

One such intermediate step is \textit{quantum error detection} (QED) codes. Unlike generic QEC codes which \textit{correct} errors, QED simply identifies if errors have occurred in the system but does not provide a mechanism for correction. Even still, QED codes are powerful for intermediate demonstrations of the important functionality required for QEC codes: repeated stabilizer measurement, mid-circuit measurement, logical operator execution, etc. When executed on their own, QED codes can provide improvement over unencoded executions of the circuit by operating as a repeat until success (RUS) procedure and discarding runs which have non-trivial syndromes. 

\begin{figure}[!t]
    \centering
    \includegraphics[width=\linewidth]{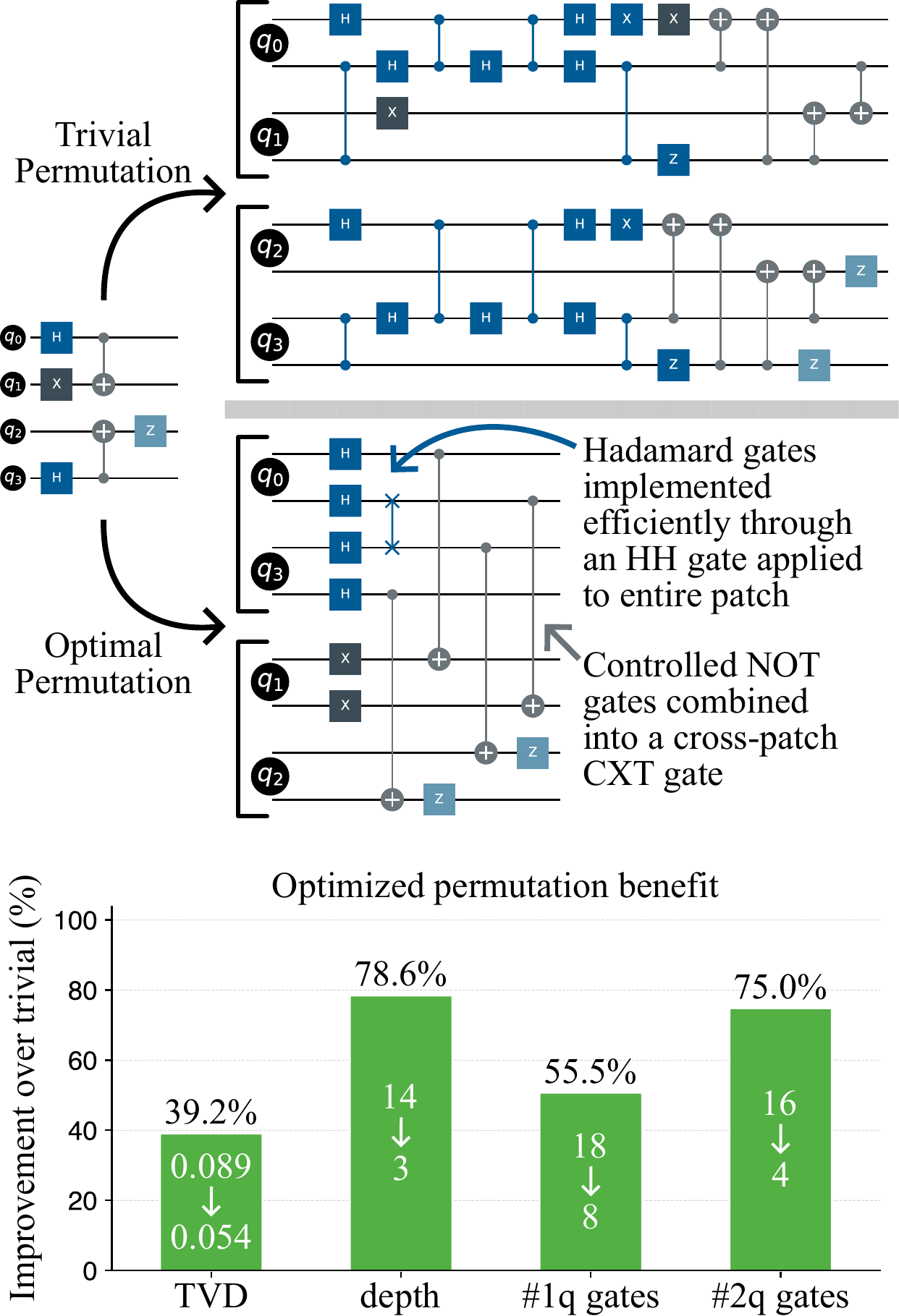}
    \caption{\textbf{Comparison of a trivial and optimized permutation for a logical circuit.} The optimal mapping enables significant gate, depth, and TVD improvements through the parallel application of two Hadamards and two CNOTs that the trivial assignment must apply individually. Swap gate not counted as discussed in Section~\ref{sec:noise_assumptions}.}
    \label{fig:complex_operations}
\end{figure}

We further stress the long-term relevance of QED codes even beyond intermediate demonstration. High distance codes are likely needed to achieve the necessary logical error rates to enable large-scale applications. One strategy is to choose a \textit{family} of codes like surface or color codes~\cite{litinski_game_2018,kubica_abcs_2018} and select an appropriate distance. Another strategy which is gaining popularity~\cite{gidney_yoked_2023,litinski_blocklet_2025}, is code concatenation which better matches modular approaches to scalability. In this construction, both the inner and outer codes can be selected independently to try to obtain the advantages of each. In this work, we focus on the Iceberg Code family~\cite{self_protecting_2024,ginsberg_quantum_2025,bedalov_fault-tolerant_2024}, which has the parameters $[[n, n-2, 2]]$, and specifically the $n = 4$ case: the $[[4, 2, 2]]$ code, the smallest code which encodes more than 1 logical qubit with detection capabilities. This code has a well-defined computational model enabling universal computation, including native gates and magic state injection, which makes it appealing as the outer code in concatenation schemes. Prior work has provided theoretical detail for how to execute circuits in this model including fault tolerance~\cite{chao_fault-tolerant_2018} and most recently on hardware for small circuits~\cite{self_protecting_2024}. Therefore, the optimization pipeline we propose is both vitally relevant for near-term demonstrations like those by Infleqtion~\cite{rines_demonstration_2025}, Quantinuum~\cite{jin_iceberg_2025}, etc., and for long-term QEC architectures, especially in the context of concatenated codes.

We address a fundamental and currently unexplored problem: efficiently compiling generic circuits and mapping program qubits onto an architecture with many Iceberg code patches. Unlike traditional logical compilation, we also need to decide which program qubits get paired and assigned to the same patch. Contrast this with the majority of FTQC architectures, e.g. surface codes or color codes, in which a single qubit is located in a single code. In these cases, the focus of compilation pipelines centers around where in the system the program qubits get mapped, typically motivated by proximity to other qubits they interacts with and to resource state production\cite{molavi2025dependency, watkins2024high, leblond2024realistic, hua2021autobraid}. These metrics are still important for codes with $k>1$ but we now have the additional choice of whether or not qubits should be located into the same encoding. In recent works which consider qLDPC codes~\cite{he_extractors_2025,sethi_optimizing_2026}, this choice has only limited effect on physical circuit execution since computation is performed on the encoded information via projective measurements. These works do not account for circuit optimizations at the physical level since they target long-term FTQC systems with sufficiently low error rates and large qubit counts. However, for the Iceberg family, we have a computational model which permits gates directly and is quite promising for near-term FTQC when we are limited in terms of qubit counts, connectivity and error rates.

In this work, we propose a complete compilation framework which compiles input programs into physical operations on Iceberg patches. Figure~\ref{fig:pipeline_flow} shows a high level overview of \acronym{}'s compilation pipeline. Specifically, \acronym{} optimizes qubit mapping to reduce \textit{gate execution} overheads between program qubits mapped into the same patches versus those mapped across different patches. For example, some mappings of qubits enable higher quality transversal gates which parallelize multi-qubit operators whereas others require the use of more expensive targeted operations. Furthermore, \acronym{} balances the tradeoff between space (number of physical qubits used) and error (the fidelity of the program). When many qubits are mapped into the same patch, we can save approximately $2\times$ on physical resource requirement. However, we are able to only detect half as many errors which reduces logical selection rate, i.e., how often we restart computation when an error is detected. This tradeoff is nuanced since some operations become cheaper when on densely packed mappings such as those seen in Figure~\ref{fig:complex_operations}. 

\noindent
\newpage
Our contributions are as follows:
\begin{enumerate}
    \item We formulate the problem of circuit optimization for high-rate codes as a qubit mapping problem followed by physical circuit optimization.
    \item We design an end-to-end open-source compilation pipeline~\cite{baker_group_ut_iceberg_compiler} that first compiles program circuits into encoded Iceberg circuits. We then propose gate merging and qubit packing algorithms to optimize these encoded circuits.
    \item Finally, we use these algorithms in conjunction with Hadamard commutation to reduce encoded circuit depth by $34\%$, one- and two-qubit gates by $31\%$ and $17\%$ respectively, improve LSR by $86\%$ and TVD by $1.75\times$ on average compared to the unoptimized circuit.
\end{enumerate}

\section{Background and Related Work}\label{sec:background}
\subsection{Quantum Error Correcting Codes}
Noisy quantum systems employ quantum error correction (QEC) as a technique to improve the fidelity of program execution \cite{terhal2015quantum}. A quantum error correcting code is denoted via its code parameters $[[n, k, d]]$, where $n$ is the number of physical qubits, $k$ is the number of logical qubits encoded by the code, and $d$ is the distance of the code, which quantifies its error tolerance. An error correcting code with distance $d$ can always correct up to $\lfloor\frac{d-1}{2}\rfloor$ independent errors, and can detect up to $d-1$ independent errors. Codes with $d = 2$ are also called quantum error detecting (QED) codes, since they can detect errors, but cannot correct them. A useful metric for QEC/QED codes is the encoding rate, which is computed as $k / n$, and this indicates the efficiency of the code. Codes with higher encoding rates are preferred since they require fewer physical qubits to implement the code. However, this usually comes at the cost of sacrificing the code distance, making this an interesting trade-off space for exploration. We discuss the stages of circuit execution for QEC codes below:

\subsubsection{Encoding Logical Qubits}
\begin{figure}[b]
    \centering
    \includegraphics[width=1\linewidth]{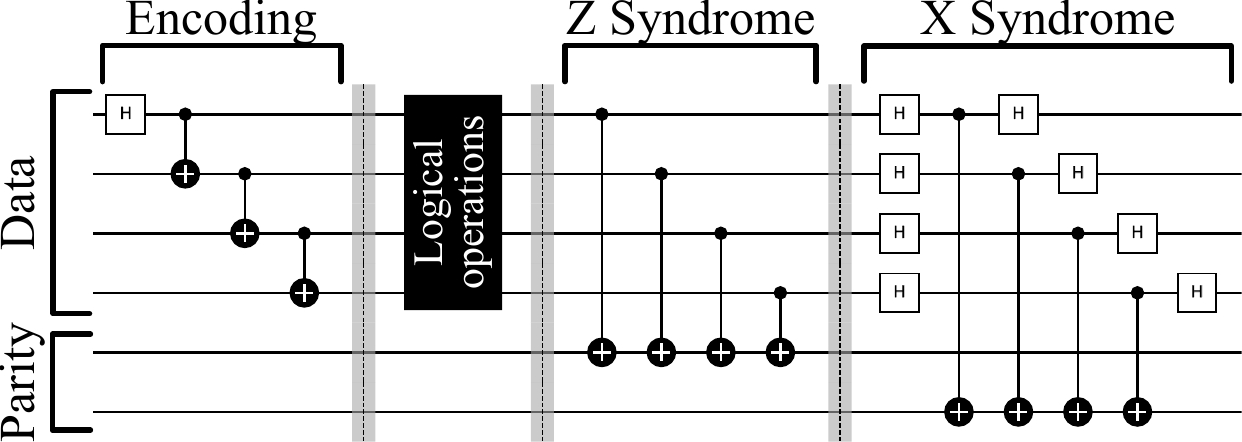}
    \caption{\textbf{Physical encoding and decoding circuit for a single Iceberg $\bm{n=4}$ patch.} Logical operations do not act on the parity bits, which are measured to extract the $Z$- and $X$-syndromes and detect $X$- and $Z$-type errors, respectively.}
    \label{fig:encoding_decoding}
\end{figure}

For a given QEC code with parameters $[[n, k, d]]$, any quantum state on $k$ qubits, $\ket{\psi} = \sum_{b \in \{0, 1\}^k} \alpha_b\ket{b}$, can be encoded into the logical state $\ket{\overline{\psi}}$ that acts on $n$ physical qubits. The circuit that performs the transformation is called the encoding circuit and is shown in Figure~\ref{fig:encoding_decoding} for the $[[4,2,2]]$ code.

\subsubsection{Memory Operations}
Once the input state $\ket{\psi}$ is encoded into the logical state $\ket{\overline{\psi}}$, the system can be maintained in the same logical state via memory operations, also commonly known as syndrome extraction circuits (Figure~\ref{fig:encoding_decoding}). These syndrome extraction circuits involve measuring additional qubits, which are called syndrome bits (or parity bits). If any of the measured syndrome bits are non-zero, this indicates an error has happened and we say this triggers an error detection event. 

\begin{figure}[t]
    \centering
    \includegraphics[width=1\linewidth]{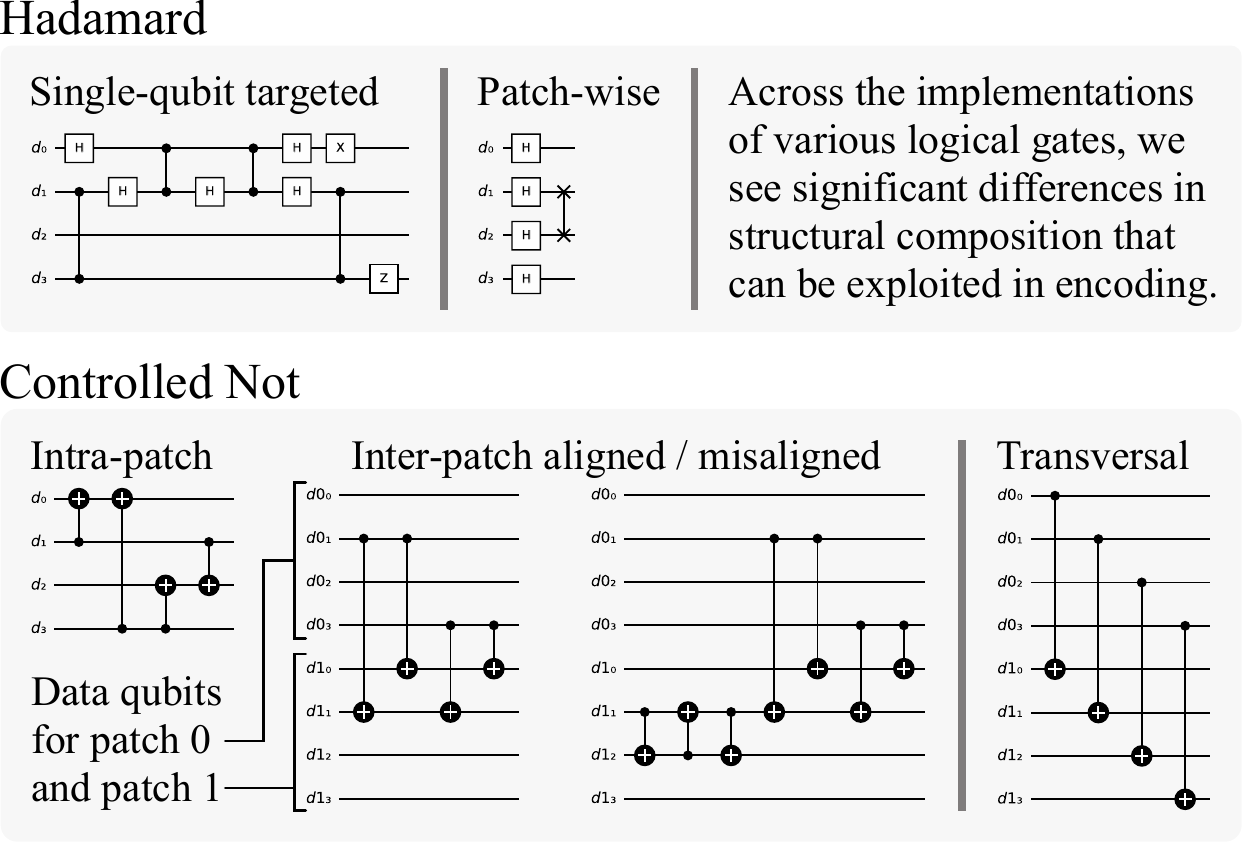}
    \caption{\textbf{Differences between Hadamard and controlled-X operations.} These differences cause significant disparities between different encodings. There are similar known differences for controlled-Z and X-controlled-X gates.}
    \label{fig:simple_operations}
\end{figure}

\subsubsection{Logical Operations}\label{sec:logical-ops}
Apart from maintaining the logical state via memory operations, we also need to be able to operate on the logical state. For every unitary $U$ that acts on the physical state and transforms $\ket{\psi} \to \ket{\phi}$, there exists a unitary $\overline{U}$ that transforms the logical state $\ket{\overline{\psi}} \to \ket{\overline{\phi}}$. Therefore, to perform universal quantum computation on the logical state, it suffices to find logical gates for a universal gate set, which is commonly the Clifford + Rz (or Clifford + T) gate set. A brute force search for the logical gates scales exponentially with the number of qubits $n$ in the code\cite{he_extractors_2025}, and there are no known efficient algorithms to compute a universal gate set for arbitrary QEC codes.


\subsection{Iceberg Codes}\label{sec:iceberg_codes}
Iceberg codes are a family of error detecting codes, and have the parameters $[[n, n-2, 2]]$. They are a promising family of QED codes since they are well-studied in literature and logical operations can be executed on the Iceberg codes with minimal overheads and trade-offs\cite{ginsberg_quantum_2025,chao_fault-tolerant_2018}. This family has a very high encoding rate $= (n-2)/n$, which approaches $1$ as the number of physical qubits is increased. However, the distance for the family is fixed at $2$, making the members of the family with higher values of $n$ less promising candidates. The most popular member of the Iceberg family is $n =4$, which is the $[[4, 2, 2]]$ code. Iceberg codes are promising for both near-term and long-term FTQC. In the near-term, when quantum computers lack resources to execute quantum error correction, we can use Iceberg codes to post-select on shot executions for which the parity bits do not trigger detection events, thereby improving the program fidelity. Additionally, in the long-term FTQC regime, concatenating Iceberg codes repeatedly can yield codes with very high distances with minimal overheads\cite{litinski_blocklet_2025}.

\subsection{Logical Operators on Iceberg Codes}\label{sec:logical_ops_Iceberg}
In Iceberg codes, a logical qubit is encoded non-locally across multiple data qubits. Consequently, logical operators are synthesized as multi-qubit physical circuits rather than single-site physical gates. The realization of these operations varies depending not only on gate type but qubit location as well as how many encoded qubits they are acting on. For that reason, logical gate cost is strongly context dependent.

This context dependence creates a spread in spacetime cost across the logical gate set known to the Iceberg code. 
Table~\ref{tab:gate_costs} summarizes the physical cost of each gate implementation, and Figure~\ref{fig:simple_operations} illustrates these differences for Hadamard and CNOT implementations. Even among single qubit gates, depth ranges from $1$ to $8$, and two-qubit gates vary by up to $4\times$ in depth depending on implementation. We exploit this disparity in \acronym{} to decide the program qubit mappings to Iceberg patches, which in turn dictates what implementations are available. For detailed implementations of the various gates see~\cite{ginsberg_quantum_2025}. Separately, we construct the X-controlled-X (XCX) from the circuit identity $H_1H_2\cdot CZ\cdot H_1H_2 = XCX$.

\begin{table}[!b]
    \centering
    \footnotesize
    \setlength\tabcolsep{0pt}
    \begin{tabular*}{\columnwidth}{@{\extracolsep{\fill}}llccc}
        \toprule
        \textbf{Gate} & \textbf{Implementation} & \textbf{1q Gates} & \textbf{2q Gates} & \textbf{Depth} \\
        \midrule
        $X$   & Targeted    & 2 & 0 & 1 \\
        $XX$  & Patch-wise  & 2 & 0 & 1 \\
        $Z$   & Targeted    & 2 & 0 & 1 \\
        $ZZ$  & Patch-wise  & 2 & 0 & 1 \\
        $S$   & Targeted    & 2 & 1 & 3 \\
        $RZ$  & Targeted    & 1 & 2 & 3 \\
        $RX$  & Targeted    & 1 & 2 & 3 \\
        $SX$  & Targeted    & 4 & 1 & 4 \\
        \midrule
        $H$   & Targeted    & 7 & 4 & 8 \\
        $HH$  & Patch-wise  & 4 & 0 & 1 \\
        \midrule
        $CNOT$
            & Intra-patch   & 0 & 4 & 4 \\
            & Inter-patch   & 0 & 4 & 3 \\
            & Transversal   & 0 & 4 & 1 \\
        \midrule
        $CZ$
            & Intra-patch   & 1 & 3 & 3 \\
            & Inter-patch   & 1 & 4 & 3 \\
            & Transversal   & 0 & 4 & 1 \\
        \midrule
        $XCX$
            & Intra-patch   & 5 & 3 & 5 \\
            & Inter-patch   & 8 & 4 & 5 \\
            & Transversal   & 8 & 4 & 3 \\
        \bottomrule
    \end{tabular*}
    \vspace{0.5em}
    \caption{\textbf{Physical cost of logical gate implementations.} $XX$ and $ZZ$ are $2\times$ as efficient as their targeted counterparts whereas $HH$ is significantly cheaper than the targeted $H$. The various gate counts and depths of the different two-qubit implementations inform bias terms discussed in Section~\ref{sec:bias_terms}.}\label{tab:gate_costs}
\end{table}

\section{The Qubit Packing Problem}\label{sec:formalizing}
Because gate implementations on the Iceberg code vary significantly depending on whether operands are targeted, patch-wise, intra-patch, inter-patch, or transversal (Section~\ref{sec:logical_ops_Iceberg}), the choice of which program qubits share a patch directly affects the error rate, resource overhead, and execution time of the compiled circuit. We refer to this choice as the \emph{qubit packing problem}. Given a program circuit with $n$ qubits and an $[[n_{code}, k, d]]$ code, the objective is to find an assignment of program qubits to patches that minimizes the encoded circuit's resource overhead and maximizes output fidelity.

For an $n$-qubit program, the number of patches, $\mathcal{P}$, can range from $\lceil \frac{n}{k} \rceil$ (dense, every patch full) to $n$ (sparse, one qubit per patch). The number of permutations between these extremes grows as $O(n!)$ which makes exhaustive search infeasible beyond a few program qubits. We define \emph{packing density} as $n / (k\times \mathcal{P})$, the ratio of program qubits to the total number of logical qubits across all patches. This metric is explored further in our evaluation (Section~\ref{sec:evaluation}).

\begin{figure}
    \centering
    \includegraphics[width=1\linewidth]{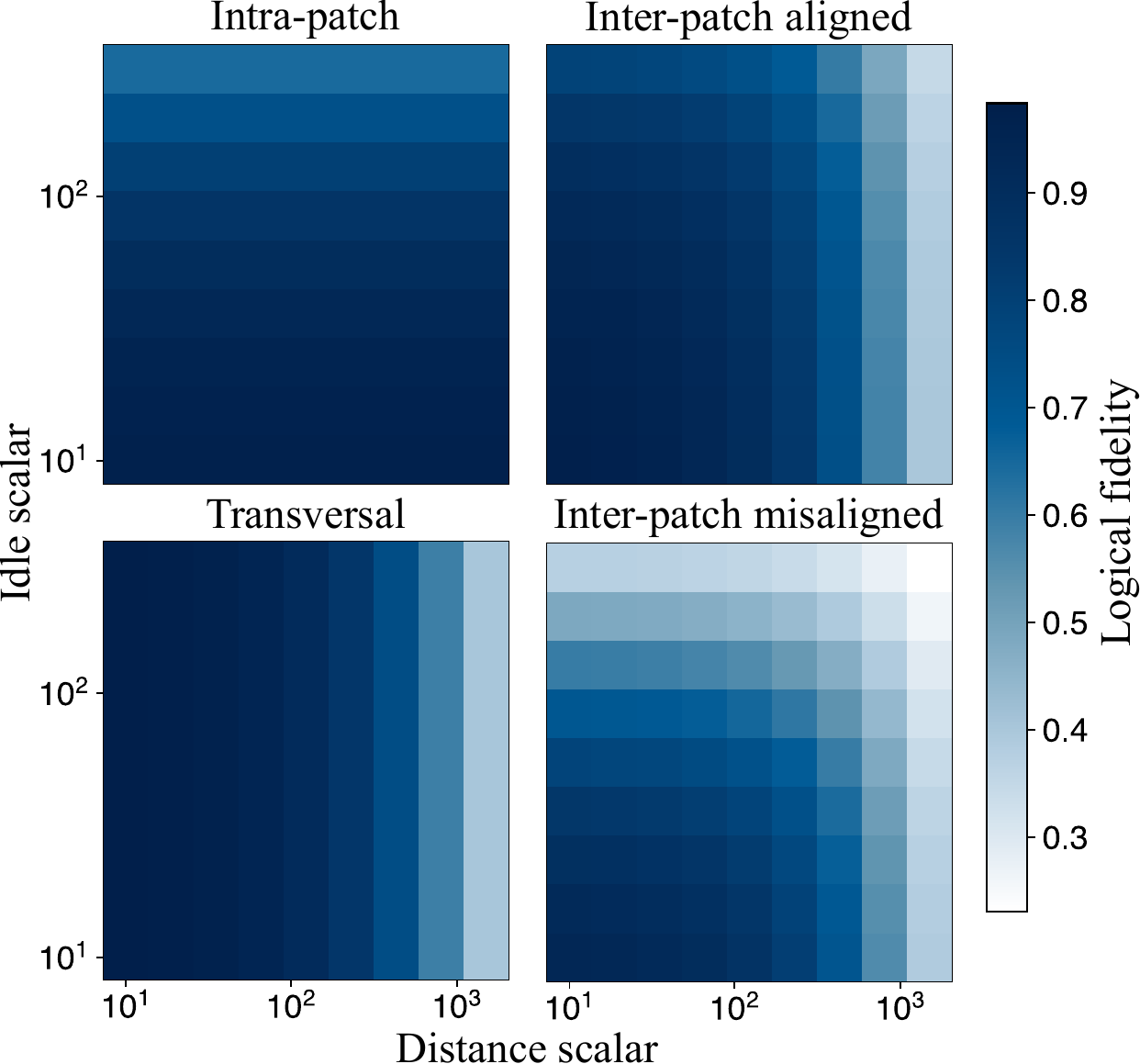}
    \caption{\textbf{Logical fidelities of different CNOT implementations (Figure~\ref{fig:simple_operations}) under varying idle- and distance-induced noise averaged across 25 random circuits and $10^5$ repetitions each.} Idle noise affects intra- and inter-patch CNOTs while distance-based noise affects inter-patch and transversal CNOTs. Implementation costs may be seen in Table~\ref{tab:gate_costs}.}
    \label{fig:cnot_heatmaps}
\end{figure}

The choice of packing density, or equivalently, how many patches to use presents a fundamental tradeoff. Fewer patches increase the opportunity for gate merging and reduce communication cost (e.g., swap insertion or atom movement), but reduces total detection capacity and exposes more logical qubits to correlated errors within each patch. Conversely, more patches improve error detection as unused logical qubits act as gauge degrees of freedom and more errors per program qubit can be detected, but increases spatial and qubit resource overheads. The optimal operating point depends on the relative strength of idle noise versus distance-based noise and the desired qubit count. Figure~\ref{fig:cnot_heatmaps} illustrates this tradeoff, showing how intra-patch, inter-patch, and transversal CNOT implementations exhibit distinct sensitivities to each noise source.

\begin{figure*}[!ht]
    \centering
    \includegraphics[width=1\linewidth]{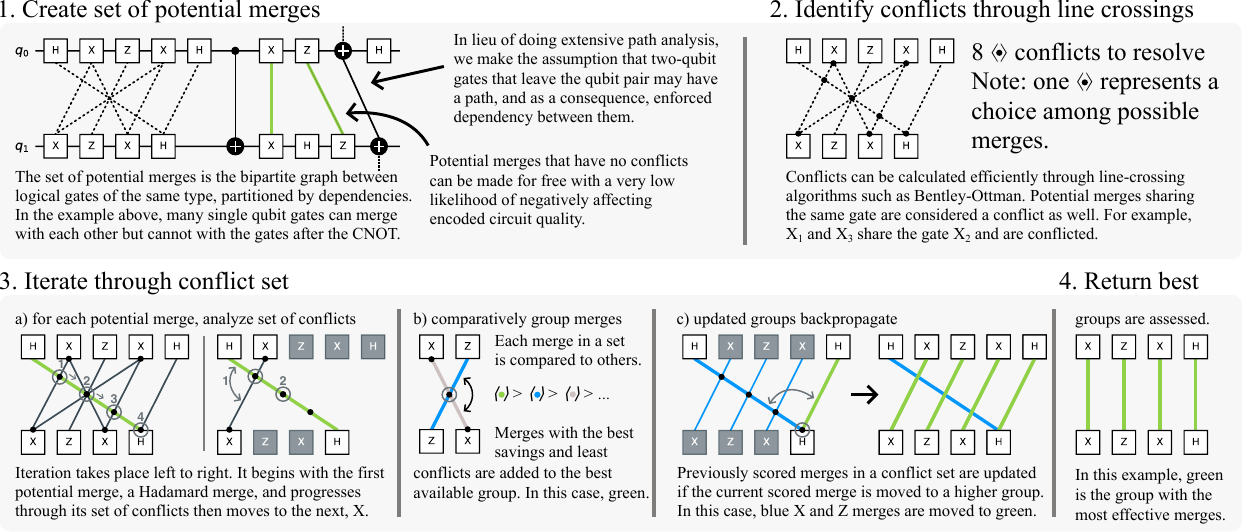}
    \caption{\textbf{Overview of the gate merging algorithm.} The algorithm runs in $O(m \cdot log(m))$ runtime, where $m$ is the circuit depth, bounded by the cost of computing line crossings. Its primary benefit is resolving conflict sets where a high-value merge pair outweighs several smaller alternatives.}
    \label{fig:merging_algo}
\end{figure*}

\subsection{The Effects of Packing}
Figure~\ref{fig:complex_operations} compares a trivial mapping with an optimized assignment for the same program circuit. In the optimized mapping, two costly single-qubit Hadamards become a single patch-wise $HH$ operation, and two sequential CNOTs collapse into a depth-one transversal gate. These substitutions improve TVD by $39.2\%$ (from 0.089 to 0.054), reduce depth by $78.6\%$ (from 14 to 3), and decrease one and two qubit gate counts by $71.4\%$ (28, 14 to 8, 4 respectively). These gains motivate qubit packing as an important optimization problem and strenthen the need to focus on permutation-aware compilation.

For small circuits, exhaustive search to find effective permutations is feasible after pruning equivalent mappings (i.e. permutations that pair the same qubits but differ only in patch ordering). However, the search space becomes intractable beyond six program qubits, exceeding billions of candidate permutations. This problem resembles dynamic bin packing where the effectiveness of each patch depends on the entire set of qubit pairings. 

\section{Our Proposal}\label{sec:proposal}
For larger circuits, the objective shifts to generating strong candidates from structure-aware compilation heuristics. Our framework utilizes Hadamard commutation, gate merging, packing, and grid alignment to produce high-performing permutations. 

\begin{figure}[!b]
    \centering
    \includegraphics[width=1\linewidth]{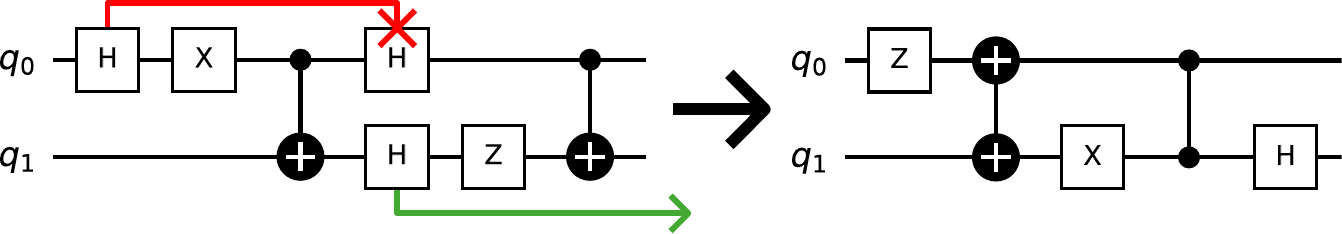}
    \caption{\textbf{Commuting Hadamards through a small circuit.} Gates Hadamards pass through experience a change in basis. Top qubit augments $X$ to $Z$ and $CX$ to $XCX$ before canceling another $H$. Bottom qubit changes an $X$ to a $Z$ and a $CZ$ to a $CX$ before reaching the end of the circuit.}
    \label{fig:h_commutation_example}
\end{figure}

\subsection{Hadamard Commutation}
Logical Hadamards on a single qubit are one of the most expensive operations on the Iceberg code (Table~\ref{tab:gate_costs}). Therefore, the first compilation pass in \acronym{} performs Hadamard commutation. After this pass, we are either left with no Hadamards (since they all cancel out), or at most $1$ Hadamard (if there were an odd number of Hadamards in the input circuit) on each logical qubit.

To do this, the optimization pass takes the program circuit and commutes each logical Hadamard through the surrounding Clifford and supported rotation gates, updating the basis of each operation it passes through. Shifting all of the Hadamards to the end of the circuit causes sequential Hadamards to cancel, leaving at most one per qubit. The reason Hadamard commutation works for the Iceberg code is because the Iceberg code has known efficient implementations for every basis-rotated logical gate. Note that this pass would fail to work for other code families for which efficient implementations of a universal gate set are not available. From our evaluations (Section~\ref{sec:evaluation}), Hadamard commutation substantially reduces physical gate overhead and exposes additional structure for the merging and packing passes that follow. Figure~\ref{fig:h_commutation_example} illustrates this on a small circuit where three logical Hadamards are reduced to one and the gates commuted through are augmented.

\subsection{Gate Merging}\label{sec:gate_merging}
The logical circuit often contains isolated single- and two-qubit gates that can be applied in parallel to entire patches or transversally between patches. Gate merging is the next compilation pass in \acronym{} and combines these compatible operations into cheaper compiled primitives, reducing physical gate count and circuit depth.

Merging may require aligning two logical operations by inserting a virtual delay on one qubit until its partner gate on the other qubit is reached. At the compiled level, this rarely increases execution time since the merged primitive is cheaper with fewer connectivity constraints than two separate targeted operations. However, not all candidate merges can be executed simultaneously as enabling one merge might disable another. Gate merging is therefore a selection problem whose objective is finding the non-conflicting subset of merges that minimizes the circuit's overall resource overhead. 

\subsubsection{Constructing Candidate Merges}
To identify candidate merges, we examine the ordered gate sequences acting on each pair of program qubits as if they were assigned to the same patch. For each pair, we construct a bipartite graph between mergeable gates (H, X, and Z), filtered by gate type and partitioned by causal dependencies between the two qubits (Step 1. in Figure~\ref{fig:merging_algo}). Two-qubit merges are more restricted as two gates can only be paired if 1. they are the same type, 2. have no intervening operations obstructing their alignment, and 3. their control and target qubits occupy different indices within their respective patches. For example, if CNOT$_1$ acts on qubit $i_A$ in patch A and $i_B$ in patch B, and CNOT$_2$ acts on $j_A$ and $j_B$ respectively, the two can be merged if $i_A \neq j_A$ and $i_B \neq j_B$.

\subsubsection{Merge Set Optimization}
Once candidate merges are constructed, we resolve conflicts among these candidates. Such conflicts are relatively rare in random circuits after Hadamard commutation but arise frequently in structured circuits. Our conflict resolution (Figure~\ref{fig:merging_algo}) operates on connected clusters of conflicting candidates for each pair of qubits. Within each cluster, we identify the conflict points, represented as line crossings and shared origins, each corresponding to a scheduling choice. We traverse the cluster linearly, comparing each conflicting pair on three criteria: resource savings, number of conflicts, and delay cost. We prioritize merges with greater savings, fewer conflicts, and lower delays. The group with the best aggregate score is chosen as the final merge set.

\subsection{Packing}\label{sec:packing}

\begin{figure}
    \centering    
    \includegraphics[width=1\linewidth]{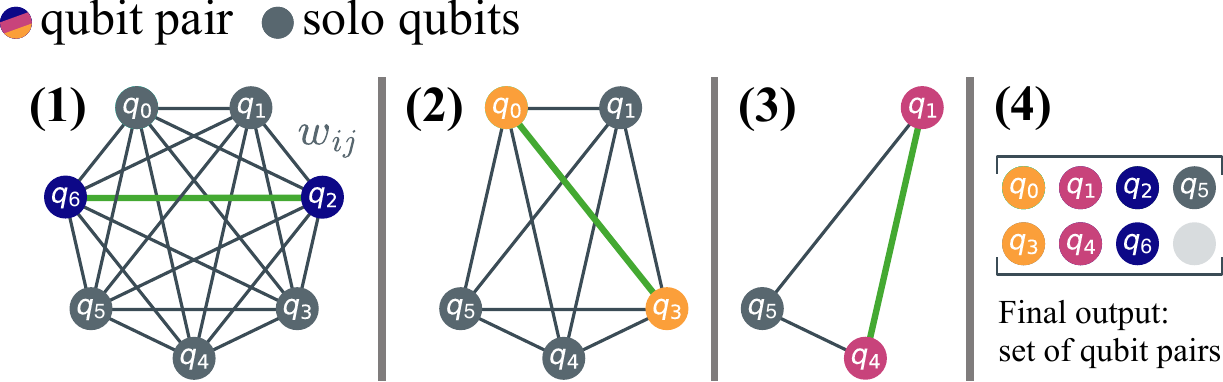}
    \caption{\textbf{Greedy assignment of qubit pairs from the score matrix.} Starting from $n$ solo qubits, each round commits the pair with the largest net gain $w_{ij}-w_{ii}-w_{jj}$ and removes both from the candidate pool. After $n - N$ rounds, remaining qubits occupy solo patches (gray); paired qubits share a color.}
    \label{fig:packing_algo}
\end{figure}

Packing determines which logical qubits should be paired, and consequently, which merge opportunities are available. While gate merging optimizes a single pair, packing optimizes the assignment of all pairs and therefore the compiled circuit as a whole. The program circuit contains fixed, causal paths of sequential operations that must occur error-free for successful execution. Merged gates shorten the compiled length of these paths, reducing physical operations and opportunities for error. It follows that packings which maximize gate merges will tend to minimize both the spacetime overhead and induced error.

\subsubsection{Qubit-Pair Scoring}
For every candidate pair of logical qubits $(i, j)$, we run the merging algorithm, described in Section~\ref{sec:gate_merging}, and record the potential savings as a tuple $\left(s^{(1)}_{ij},\, s^{(2)}_{ij}\right)$, where $s^{(1)}_{ij}$ and $s^{(2)}_{ij}$ denote the reduction in one-qubit and two-qubit physical gates respectively. These tuples populate a score matrix over all pairs.

Diagonal entries correspond to placing a qubit alone in a patch with a gauge qubit, serving as the baseline against which pairing benefits are measured. Solo qubits benefit from patch-wise Hadamards but cannot exploit two-qubit merges. Because commutation limits each qubit to at most one Hadamard, the dominant factor in solo scoring is the two-qubit gate profile. The solo penalty for two-qubit gates is computed per gate type based on the depth and gate count differences between intra-patch and transversal implementations, weighted by noise model parameters. This captures that CNOTs favor separation, CZs are more balanced, and XCXs favor pairing. Isolating qubits with fewer interactions also increases the chance that other logical qubits will have more of their gates merged.

The score matrix is symmetric up to same-patch qubit ordering, which affects qubit relabellings and does not affect circuit quality. An additional optimization pass could be inserted to determine initial qubit labels within each patch to minimize intra-patch swap gates. However, this will optimize execution only up to the first swap insertion, which yields minimal improvement. Computing the full score matrix requires $O(n^2 \cdot m \cdot \log m)$ time, where $n$ is the number of logical qubits and $m$ is the circuit depth. 

\subsubsection{Bias Terms}
\label{sec:bias_terms}
The raw merge savings $\left(s^{(1)}_{ij},\, s^{(2)}_{ij}\right)$ capture the dominant contribution in packing quality but do not distinguish between packings that yield equal merge savings yet differ in compiled depth or execution cost. To break these ties, we augment each score with bias terms that capture these secondary contributions:
\par{\emph{Depth-based delay penalty}}: Merging only requires virtual delays that typically do not affect compiled depth. However, when merged variants contain two-qubit operations that are separated by many intermediate operations, aligning them may increase compiled depth. Therefore, we add a depth-scaled penalty to merges which discourages pairings that increase execution time despite reducing gate count.
\par{\emph{Gate method preference}}: Two-qubit gate cost varies by implementation with the transversal implementation being the cheapest at depth $1$ and roughly half the physical gates of their intra- or inter-patch counterparts (Table~\ref{tab:gate_costs}). For each pair of program qubits, we compute a per-gate-type bias derived from the depth and gate count differences between intra-patch and transversal implementations, weighted by noise parameters. This favors packings that expose more transversal opportunities and applies downward pressure on circuit depth when merge counts are otherwise equal.

Concretely, the composite score for a pair $(i, j)$ is:
\[
    w_{ij} \;=\; \alpha \cdot s^{(1)}_{ij} \;+\; \beta \cdot s^{(2)}_{ij} \;-\; \gamma \cdot \Delta_{ij} \;+\; \delta \cdot T_{ij}
\]
where $\Delta_{ij}$ is the aggregate delay penalty, $T_{ij}$ is the gate-type preference bias, and $\alpha$, $\beta$, $\gamma$, $\delta$ are scaling coefficients that allow tuning between packing for idle- and distance-based noise while keeping merge savings as the dominant factor. In our experiments, we use $\alpha=\beta=1.0$, $\gamma=0.2$, and $\delta=-0.1$ to bias for depth reduction.

\begin{figure*}[!h]
    \centering    
    \includegraphics[width=1\linewidth]{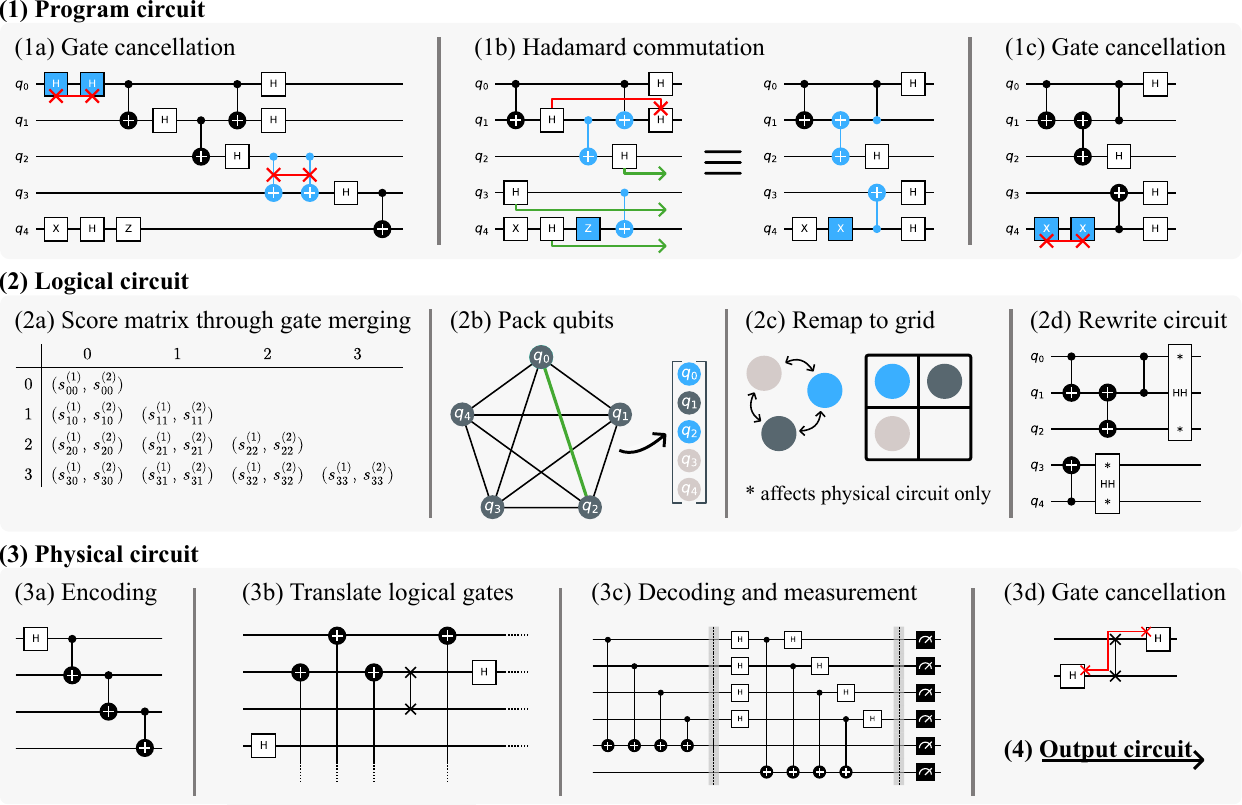}
    \caption{\textbf{Compilation pipeline for an example logical circuit.} (1a) Gate cancellation, (1b) Hadamard commutation, (1c) second gate cancellation. (2a) Pairwise merge and solo scoring, (2b) patch assignment via packing, (2c) grid alignment, (2d) circuit rewrite with merged gate macros. (3a) Patch encoding, (3b) logical-to-physical gate translation, (3c) decoding and measurement, (3d) final gate cancellation including through swap gates.}
    \label{fig:pipeline}
\end{figure*}

\subsubsection{Pair Selection}
Once composite scores are computed, packing proceeds through greedy selection parameterized by a target patch count $N$. Given $n$ program qubits, the permutation consists of $n - N$ paired patches and $2N - n$ solo patches. The procedure operates in $n - N$ rounds, outlined by Figure~\ref{fig:packing_algo}:

\begin{enumerate}
    \item \textbf{Initialize:} Assign all $n$ logical qubits to solo patches. Compute the composite score $w_{ij}$ for all candidate pairs and solo patches in the score matrix.
    
    \item \textbf{Pairing:} During each round, identify the pair $(i,j)$ with the highest score $w_{ij}$ after removing the scores of their solo patches: $w_{ij} - w_i - w_j$. Commit this pair and remove their logical qubits from the solo pool. The total packing score at round $p$ is:
    \begin{align}
        S_p(i,j) = \underbrace{\sum_{(a,b) \in \mathcal{C}} w_{ab}}_{\text{committed pairs}} + \; w_{ij} + \underbrace{\sum_{k \in \mathcal{R} \setminus \{i,j\}} w_{kk}}_{\text{remaining solos}}
        \label{eq:packing_score}
    \end{align}
    
    \item \textbf{Halt:} Repeat until $n - N$ pairs have been found. The remaining $2N - n$ qubits each occupy their own patch.
\end{enumerate}

Without a target patch count, the problem reduces to maximum weight perfect bipartite matching on a graph over two copies of the qubit set $\{0, \dots, n - 1\}$, where edge $(i,j)$ carries weight $w_{ij}$ and self-edges carry the solo score $w_{ii}$, solvable in $O(n^3)$ time. The greedy selection examines $O(n^2)$ pairs per round across $O(n)$ rounds, yielding $O(n^3)$ runtime. Combined with the $O(n^2 \cdot m \cdot \log m)$ scoring phase, packing remains tractable for circuits with thousands of logical qubits. For codes with $k>2$, pairwise scoring generalizes to $k$-subset evaluation, increasing candidate groups from $O(n^2)$ to $O(n^k)$, which may require additional pruning.

\subsubsection{Grid Alignment}
After packing, we map patches onto a two-dimensional grid to reduce inter-patch communication overhead outlined as follows:
\begin{enumerate}
    \item \textbf{Interaction Scoring:} Build an interaction graph between patches where each edge reflects the total number of two-qubit physical operations between two patches.
    \item \textbf{Linear Ordering:} Starting from the patch with the highest total interaction weight, greedily append the patch with the most interactions to the already-placed set.
    \item \textbf{Grid Placement:} Map the linear order onto a $\lceil \sqrt{\mathcal{P}} \rceil \times \lceil \sqrt{\mathcal{P}} \rceil$ grid in row-major order, where $\mathcal{P}$ is the number of patches.
\end{enumerate}
This step runs in $O(g+p^2)$ where $g$ is the number of gates and $p$ is the number of patches. More sophisticated placement algorithms exist~\cite{viszlai_matching_2025} but are beyond the scope of this work.

\subsection{Translating Logical Circuits}
The translator rewrites each logical gate into its physical Iceberg implementation, selecting the cheapest encoding based on the qubit placement: intra-patch for qubits in the same patch, inter-patch for qubits in different patches, and transversal when applicable. Because relabelings introduced during translation shift qubit positions, the translator tracks the positions throughout. The resulting compiled circuit begins and ends with the encoding and decoding sub-circuits. Figure~\ref{fig:pipeline} summarizes the full compilation pipeline.

\label{sec:evaluation}
\begin{figure*}[!ht]
    \centering
    \includegraphics[width=.95\linewidth]{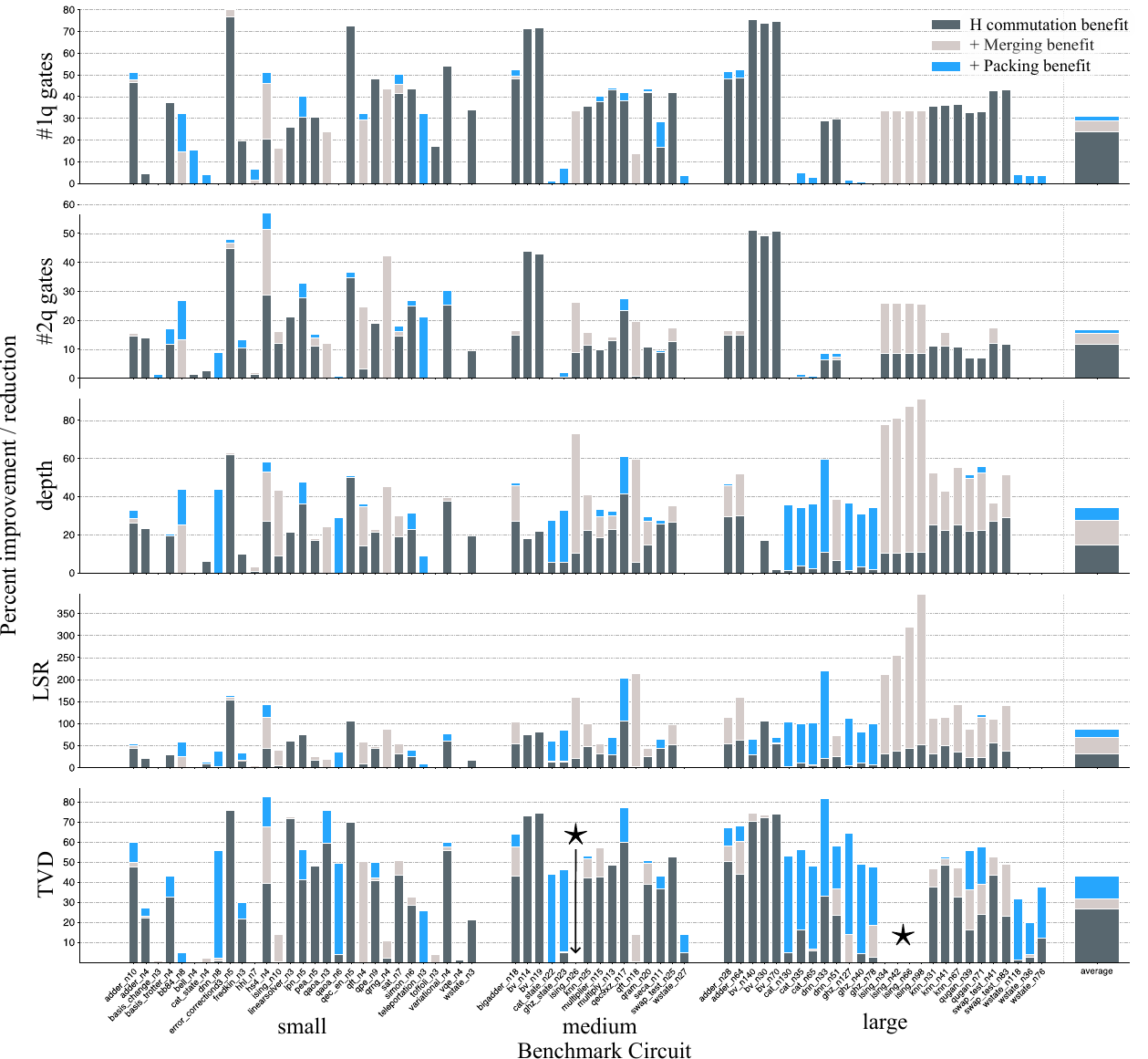}
    \caption{
    Our framework enables significant resource reduction for 71 QASMBench circuits. Across these benchmarks, \acronym{} reduces circuit depth by $34\%$, gate counts by $31\%$, $17\%$ for one and two qubit gates respectively, logical selection rate by $86\%$, and total variation distance by $43\%$ on average. We use $\lceil n/2\rceil$ patches and compare against a baseline that packs logical qubits in program order. Most benchmark circuits benefit significantly from Hadamard commutation. Circuits that have significant symmetry can exploit parallelism and benefit from merging and packing. $\star$: the Ising circuits creates large superpositions which requires exponentially scaling shots to estimate improvements in TVD which becomes intractable for $n>10$.}
    \label{fig:spacetime_improvement}
\end{figure*}

\section{Evaluation}

\subsection{Experimental Setup}
Our framework targets the $[[4,2,2]]$ Iceberg code as the most fundamental example of $k>1$ logical compilation. Because this code detects but cannot correct errors, we use a single round of error detection: the entire encoded circuit is executed, syndrome bits are measured once at the end, and shots flagging a detection event are post-selected away. Multiple rounds could be used to improve fidelity, but add complexity in terms of time and gates; a single round suffices to evaluate whether compilation techniques meaningfully reduce error rates and resource overhead. Simulations are performed using Stim\cite{gidney2021stim} by constructing Clifford proxies of the circuit~\cite{das_imitation_2023}, enabling efficient analysis of circuits exceeding ten logical qubits. We adopt a simplified noise model with the following characteristics:
\begin{enumerate}
    \item One- and two-qubit gates incur depolarizing noise approximated through Pauli twirling, parameterized by per-gate error rates $p_1$ and $p_2$, relaxation and dephasing times $T_1$ and $T_2$, and measurement error rate $p_m$.
    \item Qubits accumulate idling noise proportional to their inactive duration.
    \item Inter-patch two-qubit gates incur additional movement noise scaling with patch distance on the grid, while intra-patch operations require no movement assuming hyper-optimized operations.
\end{enumerate}
This noise model captures the three dominant error sources in encoded circuits: gate infidelity, idle decoherence, and communication cost. Due to variance in benchmark circuit size, the base noise parameters in Table~\ref{tab:noise_params} are scaled with circuit depth and the number of repetitions is adjusted with qubit count to target a post-selection rate of $\sim0.2$. These scaling choices are held constant across all techniques for each benchmark to ensure relative performance differences are attributable to the compilation.

\subsubsection{Evaluation Metrics}
Static spacetime metrics such as gate count and depth do not fully capture mapping quality because they do not reflect error propagation under noise. We additionally evaluate mappings using total variation distance (TVD), which measures how close the output distribution is to a reference, and logical selection rate (LSR), which measures what fraction of shots pass post-selection. Explicitly, we define them as:
\[
\mathrm{TVD}\!\left(P_{\mathrm{noisy}}^{\mathrm{post}}, P_{\mathrm{ideal}}^{\mathrm{post}}\right)
=
\frac{1}{2}
\sum_{x}
\left|
P_{\mathrm{noisy}}^{\mathrm{post}}(x)
-
P_{\mathrm{ideal}}^{\mathrm{post}}(x)
\right|;\quad
\mathrm{LSR} = \frac{M_{\mathrm{kept}}}{M_{\mathrm{total}}},
\]
where $P_\mathrm{noisy}^\mathrm{post}, P_\mathrm{ideal}^\mathrm{post}$ are the probability distributions obtained after post-selecting on samples for which none of the syndromes fired for any of the Iceberg codes, for the noisy and ideal simulations, respectively. $M_\mathrm{kept}, M_\mathrm{total}$ are the number of samples that are post-selected, and the total number of samples, respectively.

\subsubsection{Simulation Scope and Noise Assumptions}
\label{sec:noise_assumptions}
We assume each logical circuit has been precompiled and baseline-optimized; we do not re-optimize gate synthesis per mapping. We additionally assume intra-patch swap gates can be implemented as a qubit relabeling at negligible cost. This holds for architectures with flexible connectivity such as neutral atom systems but may require revision for systems with fixed topologies. Table~\ref{tab:noise_params} lists the base parameters we use for the simulator.
\begin{table}[!h]
    \centering
    \footnotesize
    \setlength\tabcolsep{0pt}
    \caption{\textbf{Stim noise model parameters.} We scale these values depending on the experiment as discussed in Section~\ref{sec:evaluation}.}
    \label{tab:noise_params}
    \begin{tabular*}{\columnwidth}{@{\extracolsep{\fill}}cll}
        \toprule
        \textbf{Parameter} & \textbf{Value} & \textbf{Description} \\
        \midrule
        $p_1$     & $1 \times 10^{-3}$  & 1-qubit gate error probability \\
        $p_2$     & $5 \times 10^{-3}$  & 2-qubit gate error probability \\
        $p_m$     & $2 \times 10^{-2}$  & measurement error probability \\
        $T_1$     & $100\,\mu\text{s}$  & amplitude damping time \\
        $T_2$     & $80\,\mu\text{s}$   & dephasing time \\
        $t_{1q}$  & $50\,\text{ns}$     & 1-qubit gate duration \\
        $t_{2q}$  & $300\,\text{ns}$    & 2-qubit gate duration \\
        \bottomrule
    \end{tabular*}
\end{table}

\begin{figure*}
    \centering
    \begin{subfigure}{.9\linewidth}
        \centering
        \includegraphics[width=\linewidth]{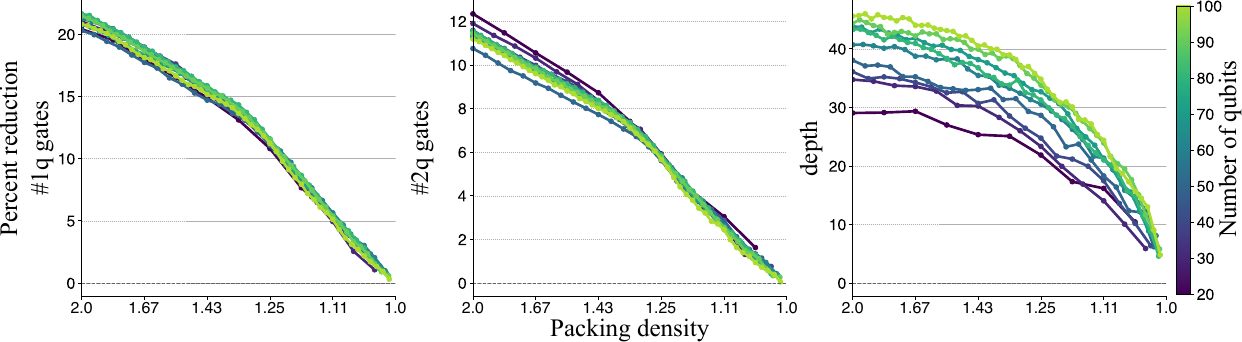}
        \caption{Fixed circuit depth (10 gates per qubit), varying qubit count.}
        \label{fig:random_sweep_qubits}
    \end{subfigure}
    \begin{subfigure}{.9\linewidth}
        \centering
        \includegraphics[width=\linewidth]{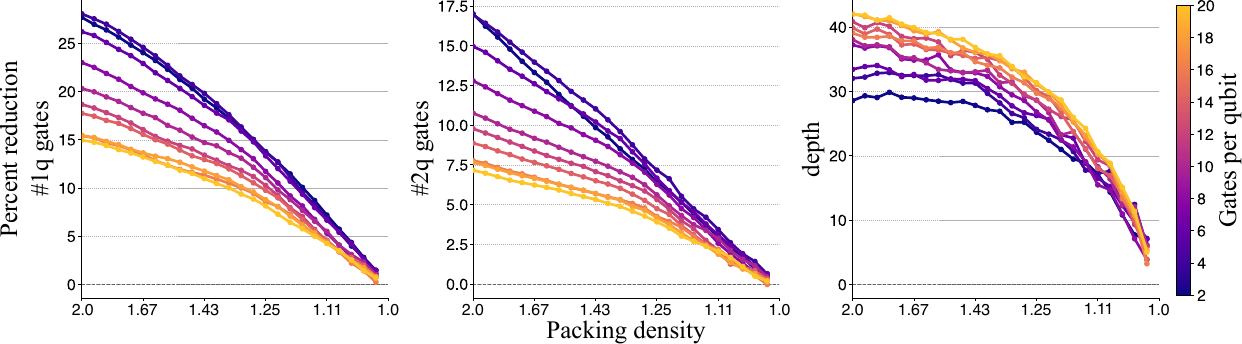}
        \caption{Fixed qubit count (50), varying gates per qubit.}
        \label{fig:random_sweep_gates}
    \end{subfigure}
    \caption{\textbf{Packing and merging improvement over random permutations by packing density.} Each point compares \acronym{}'s permutation against the mean of $40$ random  permutations at equal patch count, averaged over $20$ random circuits. Reductions are greatest at density $2.0$ (all patches full) and converge to zero at density $1.0$ (all solo patches). (a)~Fixed depth ($10$ gates per qubit), varying qubit count $20$--$100$. Gate reductions decline linearly with a steepening slope below $\sim 0.78$; depth follows a concave trajectory, maintaining most gains through moderate densities. Larger circuits benefit more from a richer candidate pool. (b)~Fixed qubit count ($50$), varying depth $2$--$20$ gates per qubit. Shallower circuits see greater gate reductions; deeper circuits see greater depth reductions but also more merge conflicts.}
    \label{fig:random_sweep_combined}
\end{figure*}

\subsection{Spacetime Resource Reduction}
\label{sec:spacetime_reduc}

\sloppy We evaluate \acronym{} on a suite of 71 benchmark circuits taken from QASMBench~\cite{li2022qasmbenchlowlevelqasmbenchmark}, selecting all circuits up to 150 logical qubits and 2,500 gates after transpiling to the gate set $\{H, X, Z, CX, CZ, S, R_x, R_z\}$. Non-Clifford circuits are simulated via proxy circuits that replace rotations with phase gates. This preserves relative performance differences between proxy and actual because phase and arbitrary rotations incur similar error profiles. These benchmarks span a wide range of state preparation, arithmetic, variational algorithms, and error correction circuits. 

Across these circuits, \acronym{} achieves average reductions of $31\%$ in single-qubit gates, $17\%$ in two-qubit gates, and $35\%$ in circuit depth relative to a naive baseline that assigns qubits to patches sequentially by index with no compilation techniques applied. Figure~\ref{fig:spacetime_improvement} shows the contributions of each technique grouped by size. 

Hadamard commutation provides the majority of gate reductions, with merging and packing contributing an additional $5$--$10\%$. Depth reduction benefits more from merging and packing, since the variation among the transversal, intra-, and inter-patch implementations gives \acronym{} more room to exploit circuit structure. If the naive ordering does not expose these opportunities, the packing algorithm identifies permutations that do. 

The ordering of compilation passes affects how credit is attributed among techniques. Because Hadamard commutation is applied first, it captures the largest share of gate reductions and leaves fewer opportunities for merging and packing to merge Hadamards. If merging and packing were applied before commutation, their relative contributions would increase accordingly. This ordering effect is specific to the Iceberg code, where the targeted Hadamard dominates encoding overhead. For codes where no single gate dominates, we expect merging and packing to account for a larger fraction of the total improvement.

The efficacy of each technique depends heavily on circuit structure. Four benchmarks illustrate this range. The \texttt{dnn\_n8} circuit contains no Hadamards but features many parallel CNOT operations; packing enables transversal substitutions that reduce two-qubit gates and depth. The Bernstein--Vazirani circuits (\texttt{bv\_n*}) see large gate reductions from near-complete Hadamard removal, but depth is preserved because commutation transforms CNOTs into XCX gates whose encoded implementations contain physical Hadamard layers. The Ising circuits (\texttt{ising\_n*}) are representative of highly symmetric structures. Applying commutation removes most Hadamards, and the resulting pattern allows extensive merging into transversal operations, compounding into large reductions across all metrics. The GHZ and cat-state circuits (\texttt{ghz\_state\_n*} and \texttt{cat\_state\_n*}) benefit primarily from packing that places CNOT-heavy qubit pairs in separate patches, exploiting the depth-3 inter-patch CNOT over the depth-4 intra-patch variant. 

Compared to the baseline, our compiling techniques require substantially fewer physical operations across all size categories. The gap between naive and optimized compilation remains stable as circuits scale from small to large quantities of logical qubits, suggesting that the pipeline continues to find opportunities in larger circuits.

\subsubsection{Random Circuits}
\label{sec:random_circs}
\begin{figure}
    \centering
    \includegraphics[width=1\linewidth]{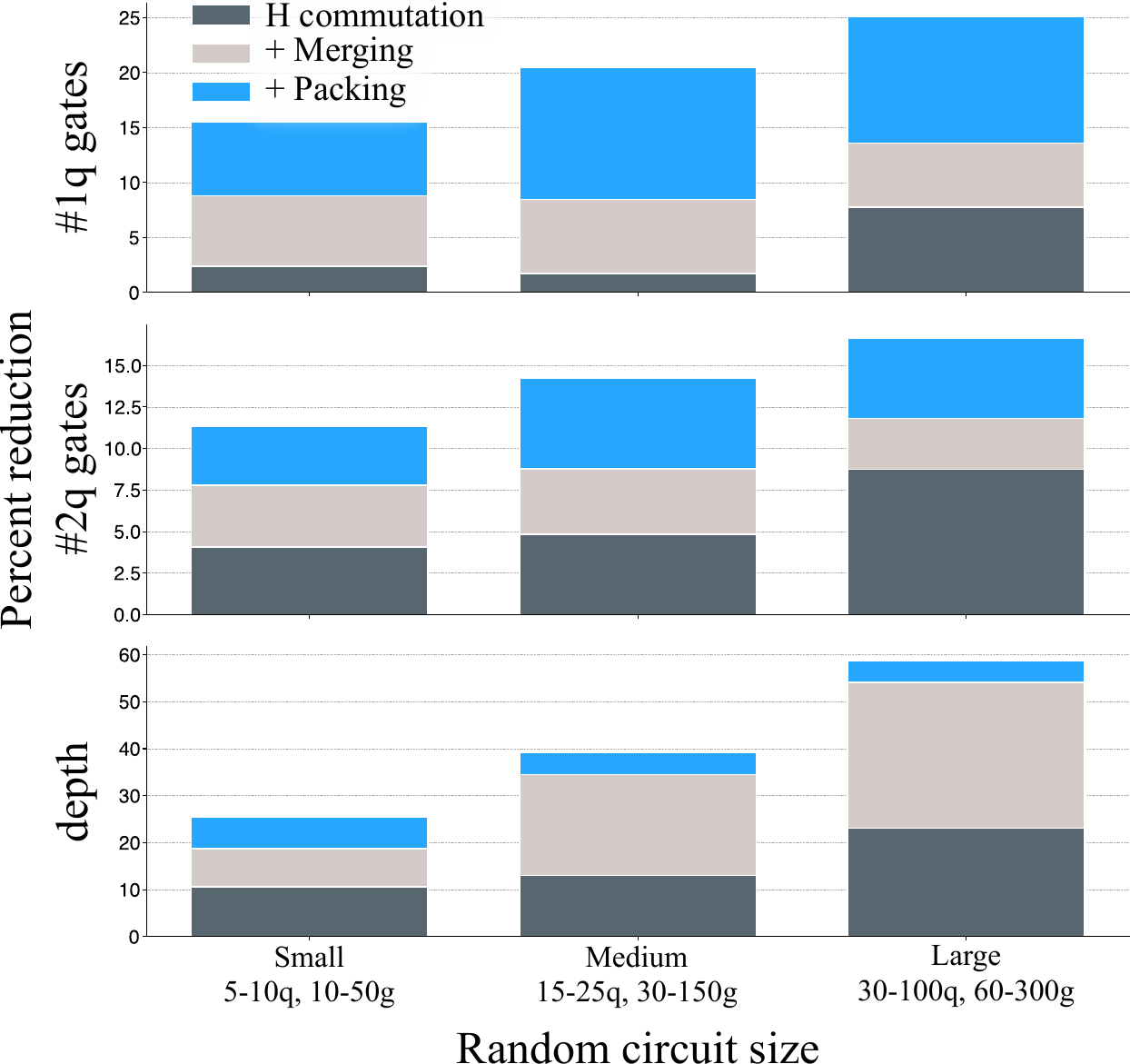}
    \caption{\textbf{Compilation techniques significantly improve resource overhead for random circuits.} Compared to benchmarks, random circuits contain fewer immutable gates ($R_x$, $R_z$, $S$), giving merging and packing greater opportunities for gate reduction. Depth reduction is comparatively smaller due to less structural symmetry and \acronym's parameterization favoring gate reduction.}
    \label{fig:spacetime_improvement_random}
\end{figure}

We observe similar trends on random circuits composed of gates from $\{X, Z, H, S, R_z, R_x, CX, CZ\}$ with a uniform any-to-any random selection and no immediate cancellations. As shown in Figure~\ref{fig:spacetime_improvement_random}, compilation benefits increase with circuit size and depth. Because random circuits contain fewer immutable gates ($S$, $R_x$, $R_z$) than the benchmark set, a larger fraction of the circuit is amenable to merging, yielding greater gate reductions. Depth sees less improvement since random circuits exhibit less structural symmetry for packing to exploit.

This contrast highlights an important aspect of our proposed techniques: the set of logical gates available to \acronym{} dictates which circuit structures may be exploited. Gates with only one known Iceberg implementation act as immutable obstacles, and rotation gates cannot be merged due to their varying parameterization. As additional implementations of encoded gates become known, the scope of logical compilation will grow accordingly. 

\subsubsection{Qubit Density and Circuit Parameters}
The results above use the tightest possible packing, where every patch holds two logical qubits. To isolate the packing heuristic's value from the general benefit of dense packing, we compare \acronym's chosen permutation against the average of $40$ random permutations at each patch count, sweeping packing density from $2.0$ (dense) to $1.0$ (sparse). Each data point averages across $20$ random circuits.

Figure~\ref{fig:random_sweep_combined} characterizes how packing density affects technique effectiveness. Gate reductions are greatest at the tightest packing and decline gradually, preserving much of the benefit even at moderate densities. Depth reduction follows a concave trajectory, maintaining most gains until nearly all qubits are placed solo. This decline reflects a gradual exit from the regime where merging and packing are relevant as $k\to1$. Conversely, we expect the efficacy of these techniques to become more significant as $k$ increases. These trends strengthen with circuit size. Additionally, because the optimal tradeoff between gate count and depth depends on whether distance-based or idle noise dominates, tuning the composite score coefficients (Section~\ref{sec:bias_terms}) to the target hardware's noise profile could yield further improvement.

\begin{figure}
    \centering    
    \includegraphics[width=1\linewidth]{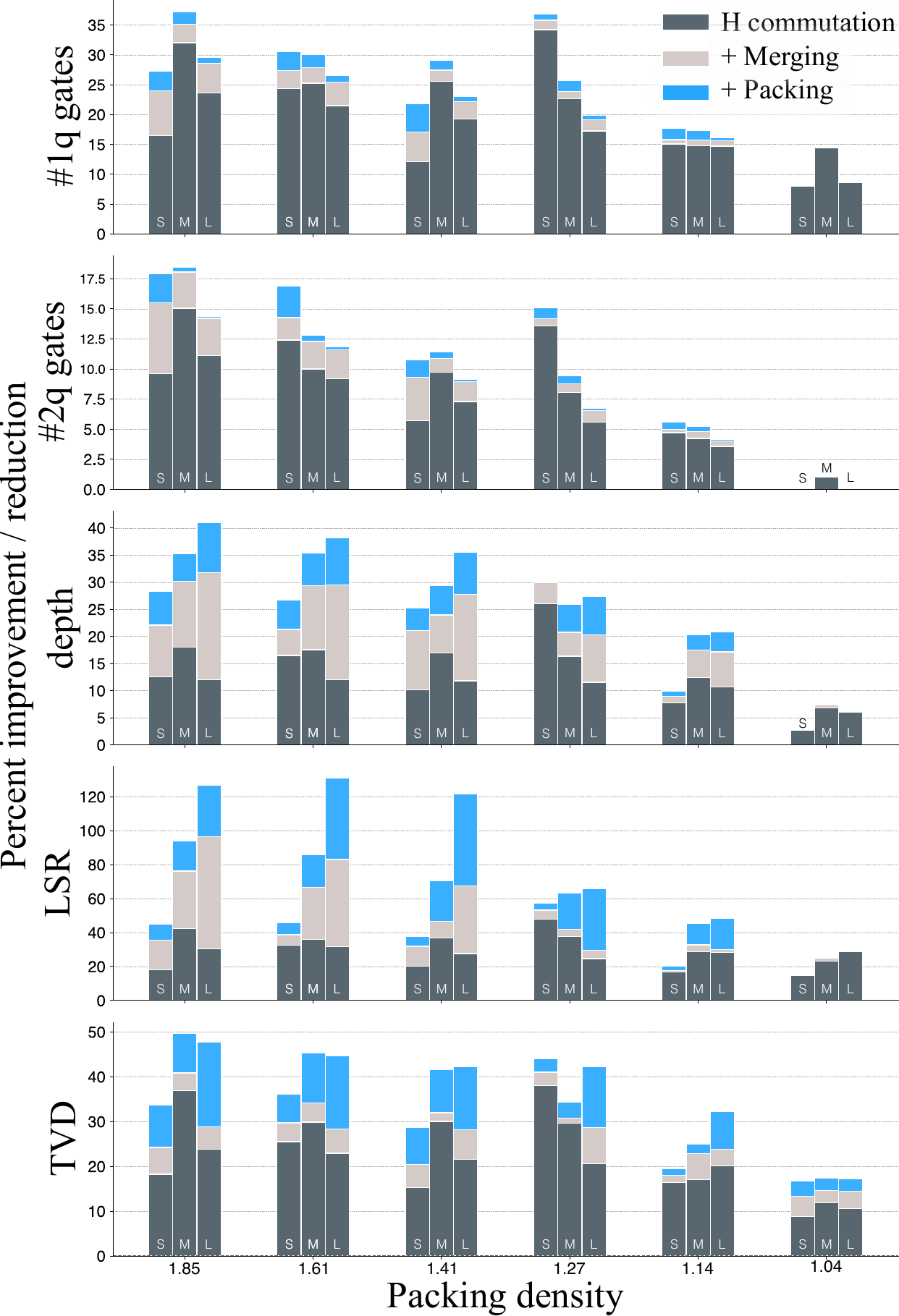}
    \caption{\textbf{Ablation of techniques over packing density.} Hadamard commutation maintains consistent benefits across all densities. Merging and packing benefits decline roughly linearly for gate reductions but follow a concave curve for depth. Notably, packing contributes disproportionately to LSR and TVD improvement even at lower densities, attributable to grid alignment reducing communication cost.}
    \label{fig:ablation_sweep}
\end{figure}

Figure~\ref{fig:ablation_sweep} shows the ablation of each technique over the same packing density sweep, averaged across small, medium, and large benchmark circuits. Hadamard commutation maintains consistent benefits across all densities as expected since it modifies the program circuit independently of packing. Merging and packing benefits decline with decreasing density, following the pattern in Figure~\ref{fig:random_sweep_combined}. Larger circuits see greater depth improvements but relatively small gate count improvements, reinforcing the scaling behavior observed above. The most notable result is that packing contributes disproportionately to LSR and TVD improvement relative to its gate reduction contribution, particularly at moderate densities. This is attributable to grid alignment, which reduces inter-patch communication cost regardless of how many qubits are paired, resulting in fewer errors and improved fidelity. At the tightest packings, the full pipeline achieves LSR and TVD improvements exceeding $100\%$ for large circuits.

\subsection{Scaling and Limitations}
There exist a few drawbacks to our proposal. Most notably, post-selection and the Iceberg code itself are near-term approaches. Because the $[[4,2,2]]$ code cannot correct errors, it is infeasible as the primary encoding for circuits with hundreds of logical qubits as deep circuits will incur runtime errors that post-select away most of the samples. However, while the Iceberg code may not scale, the compilation techniques (H commutation, merging, packing) are properties of the compiler, rather than the code. However, while it may vary from code to code, we expect the following techniques to improve execution through:
\begin{enumerate}
    \item Reducing the impact of burdensome encoded gates, such as the Iceberg Hadamard. (H commutation)
    \item Locating opportunities to apply multiple gates in parallel. (Merging)
    \item Determining mappings for qubits to create these opportunities while reducing communication and resource overhead. (Packing)
\end{enumerate}

Two obstacles limit the scope of these techniques. First, packing currently relies on pairwise scoring; for codes with $k>2$, this generalizes to $k$-subset evaluation, which scales as $O(n^k)$ and may require approaches such as network clustering on the qubit interaction graph. Second, \acronym's effectiveness is bounded by the gate types it can manipulate---arbitrary unitaries and rotations have only one known Iceberg implementation and cannot be merged.

Lastly, our evaluation is bounded by implementation constraints rather than algorithmic limits. The pipeline's theoretical complexity is $O(n^2 \cdot m \cdot \log m)$, and compilation for each benchmark completes within seconds on a MacBook M3 Max with 36 GB of RAM. Benchmark circuits were limited to 150 logical qubits and 2,500 logical gates ($\sim750$ physical qubits, $\sim10,000$ physical gates after transpilation). An optimized implementation with proper data structures and parallelism would extend this ceiling considerably.

\section{Conclusion}
\label{sec:conclusion}

We propose a compilation toolkit for high-rate quantum error detecting codes, specifically the Iceberg code, which addresses the previously unstudied problem of mapping logical qubits to multi-qubit patches. Using Hadamard commutation, gate merging, and packing, \acronym{} reduces depth by up to $34\%$ and improves LSR by $86\%$ and TVD by $1.75\times$ over a naive baseline. We have also shown that these gains become stronger, and more important, as the size of circuits increases. Our merge-aware, noise-biased heuristic algorithms identify high-performing permutations that improve encoded circuit quality through implementation cost reduction. 

While our evaluation targets the $[[4,2,2]]$ Iceberg code, we expect the compilation techniques presented in this work to generalize to other high-rate codes and continue to grow in importance with circuit size. Any code that encodes multiple logical qubits per patch will benefit from reducing the presence of its most expensive gates, finding opportunities for parallel gate application, and mapping logical qubits to their most effective blocks. Extending the merging and packing algorithms to $k>2$, broadening gate vocabulary to include arbitrary rotations, and optimizing compilation runtime are logical next steps. As encoded gate vocabularies and circuit sizes increase, the effect of structure-aware logical compilation will continue to grow.

Our compilation toolkit is available as open-source software at the Iceberg Compiler GitHub repository~\cite{baker_group_ut_iceberg_compiler}.

\newpage
\bibliographystyle{ref_format}
\bibliography{refs}

\appendix
\section{Appendix}
\begin{figure}
    \centering
    \includegraphics[width=.85\linewidth]{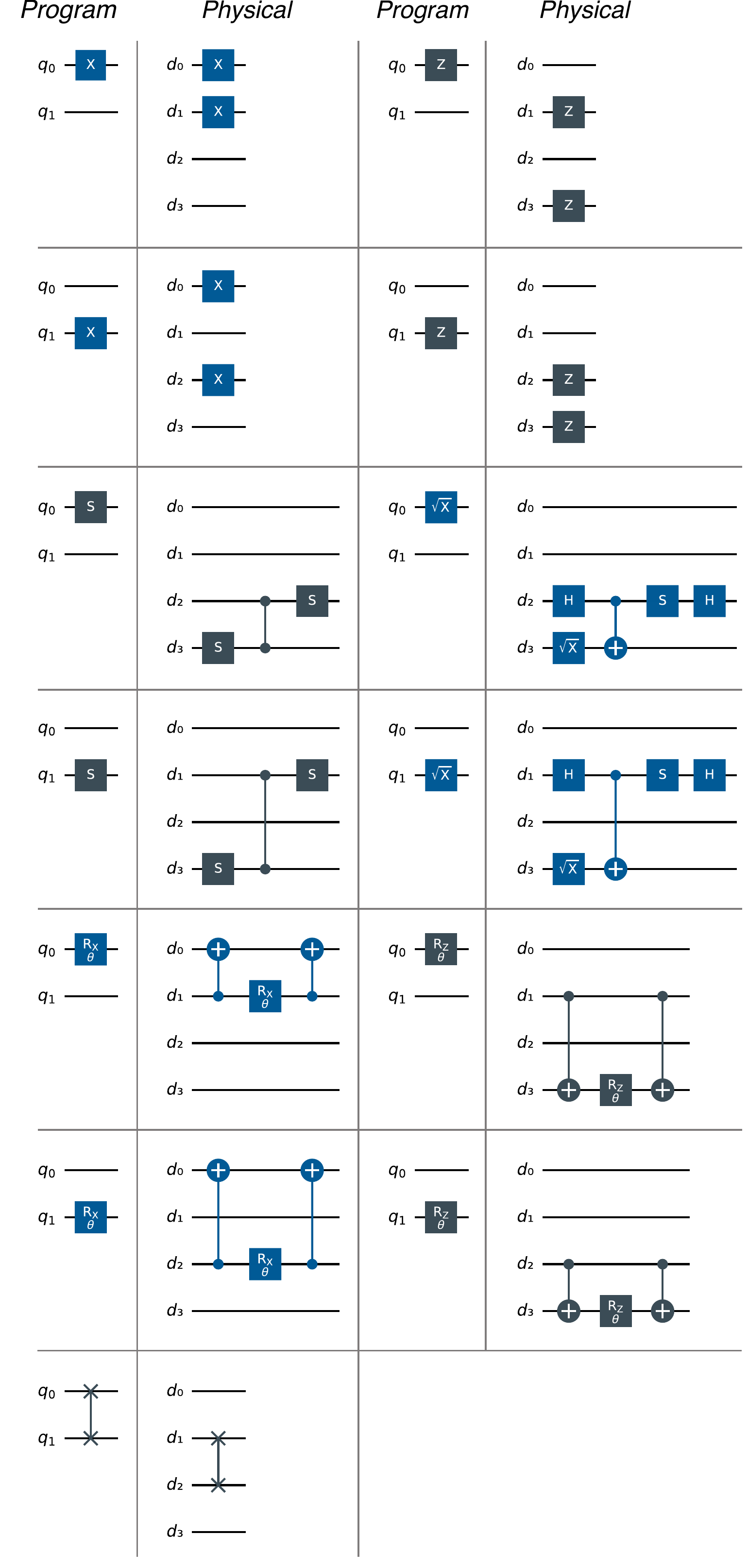}
    \caption{\textbf{Single qubit gate operations and swap} Columns two and four are the encoded versions of the gates to their left.}
    \label{fig:simple_operations_1}
\end{figure}

\begin{figure*}
    \centering
    \includegraphics[width=.85\linewidth]{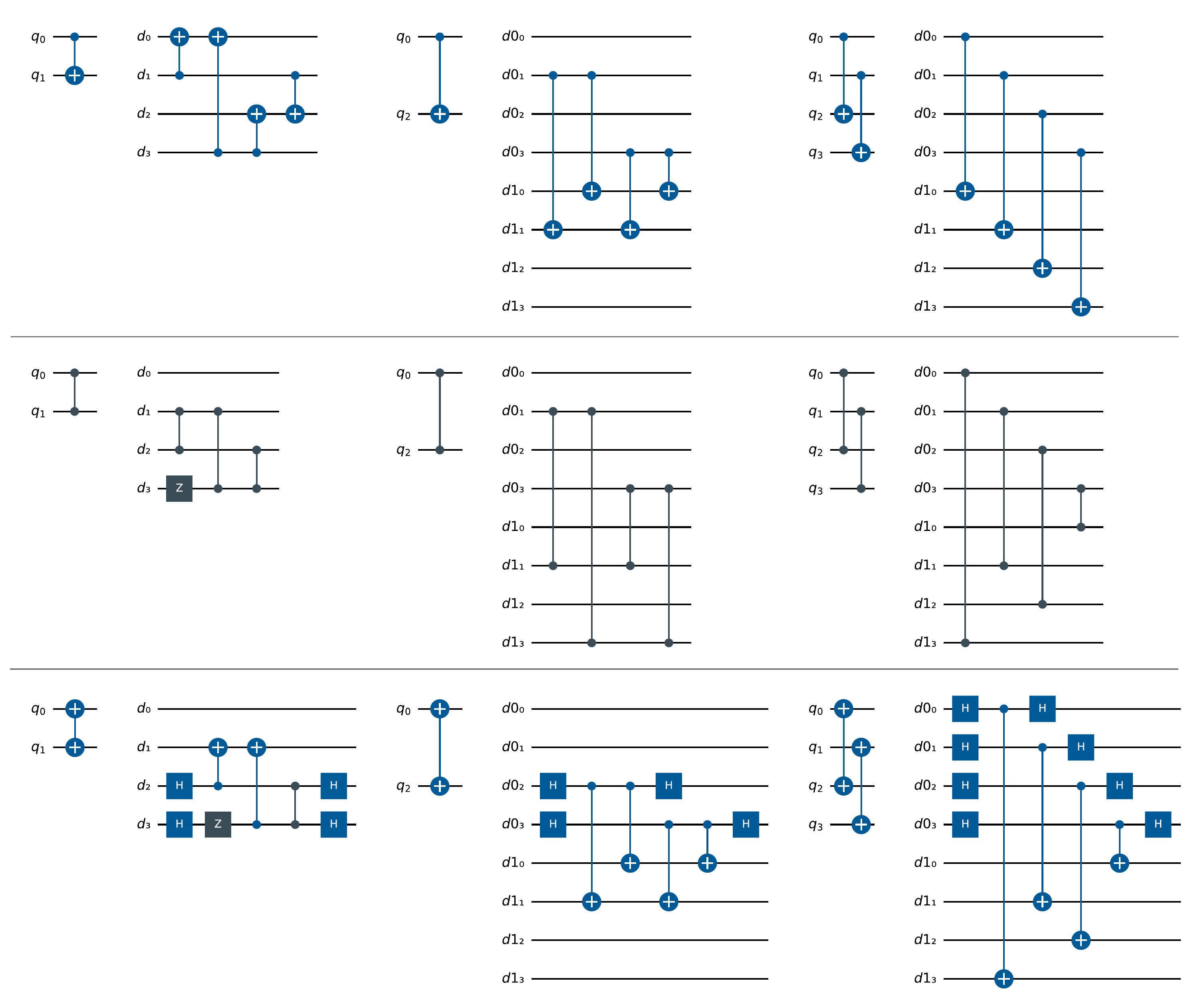}
    \caption{\textbf{CX, CZ, and XCX gate operations.} The first column is the program circuit. The second column is the intra-patch encoding. The third column is the inter-patch encoding. The final column is the transversal encoding.}
    \label{fig:simple_operations_3}
\end{figure*}

\begin{figure}
    \centering
    \includegraphics[width=.85\linewidth]{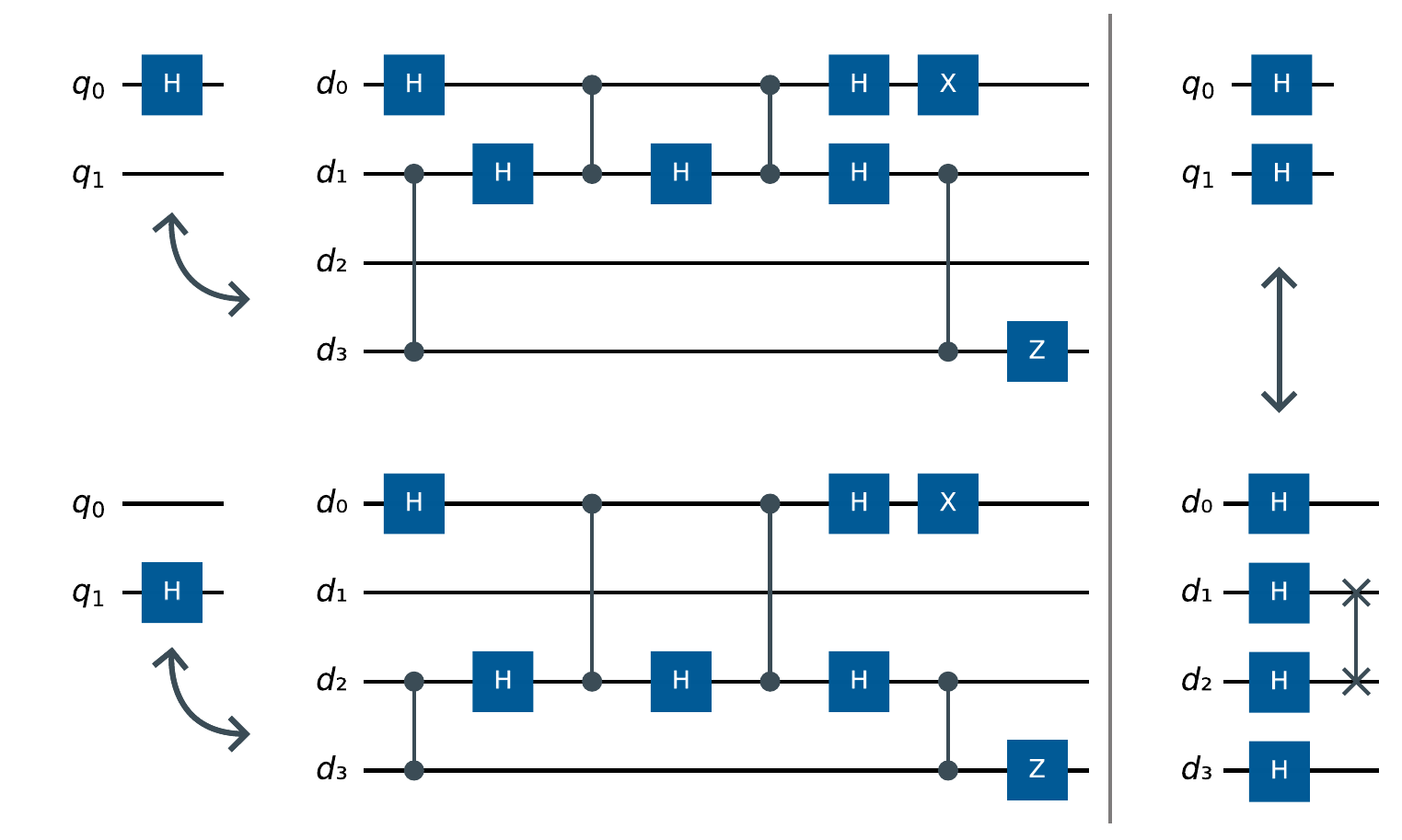}
    \caption{\textbf{Encoded Hadamard operations.}}
    \label{fig:simple_operations_2}
\end{figure}

We utilize the following gates seen in Figures \ref{fig:simple_operations_1}, \ref{fig:simple_operations_2}, \ref{fig:simple_operations_3} in the translation pass for creating encoded circuits from input program circuits. 

\subsection{Fidelity Improvement}
Reducing spacetime resources is a necessary but incomplete objective as compilation techniques must also translate circuits to improved output quality under realistic noise. In this section, we establish that logical selection rate (LSR) and total variation distance (TVD) are strongly coupled under post-selected error detection, explain why, and then show that \acronym's compilation pipeline consistently improves both metrics. 

\subsection{Selection Rate and Fidelity}
\begin{figure*}
    \centering
    \begin{subfigure}{0.48\linewidth}
        \centering
        \includegraphics[width=\linewidth]{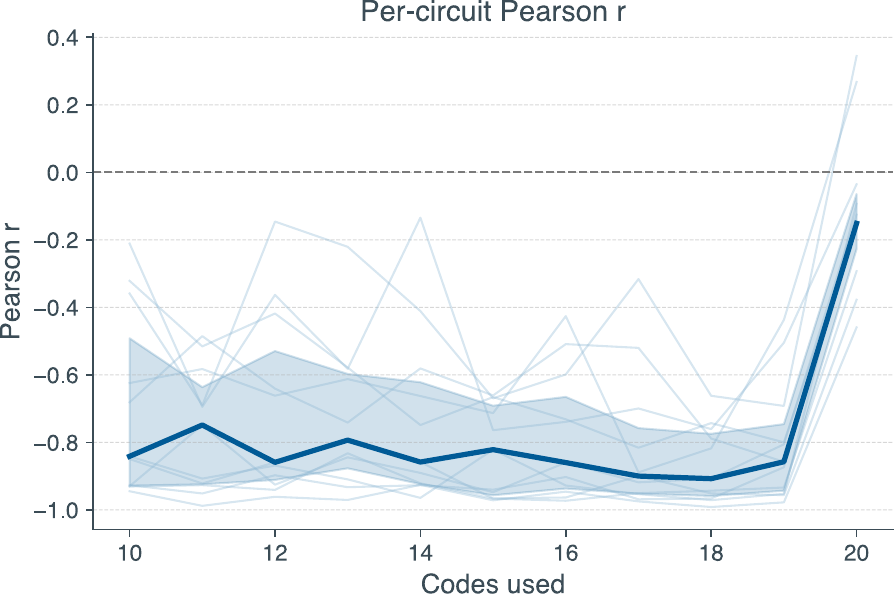}
        \caption{Patch count}\label{fig:lsr_tvd_codes}
    \end{subfigure}
    \hfill
    \begin{subfigure}{0.48\linewidth}
        \centering
        \includegraphics[width=\linewidth]{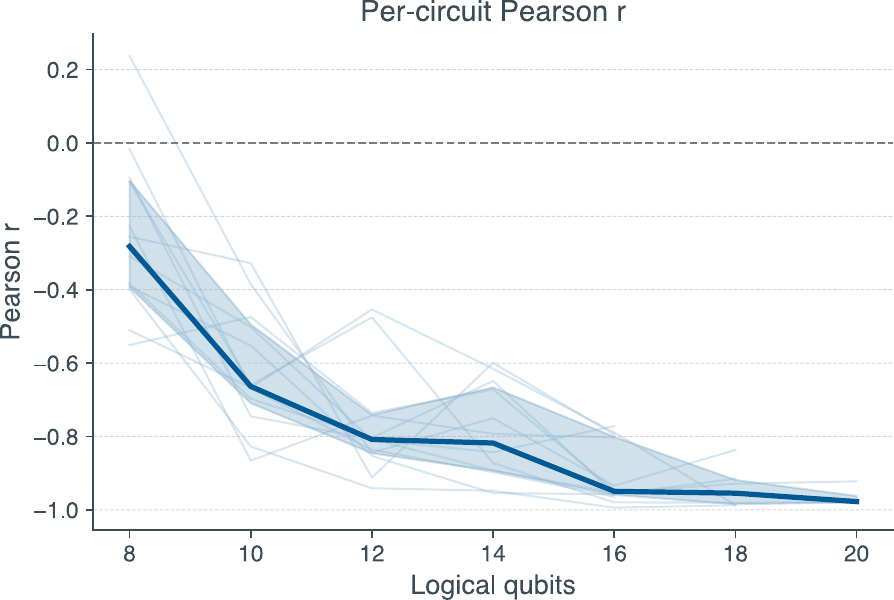}
        \caption{Qubit count}\label{fig:lsr_tvd_qubits}
    \end{subfigure}
    \vspace{0.5em}
    \begin{subfigure}{0.48\linewidth}
        \centering
        \includegraphics[width=\linewidth]{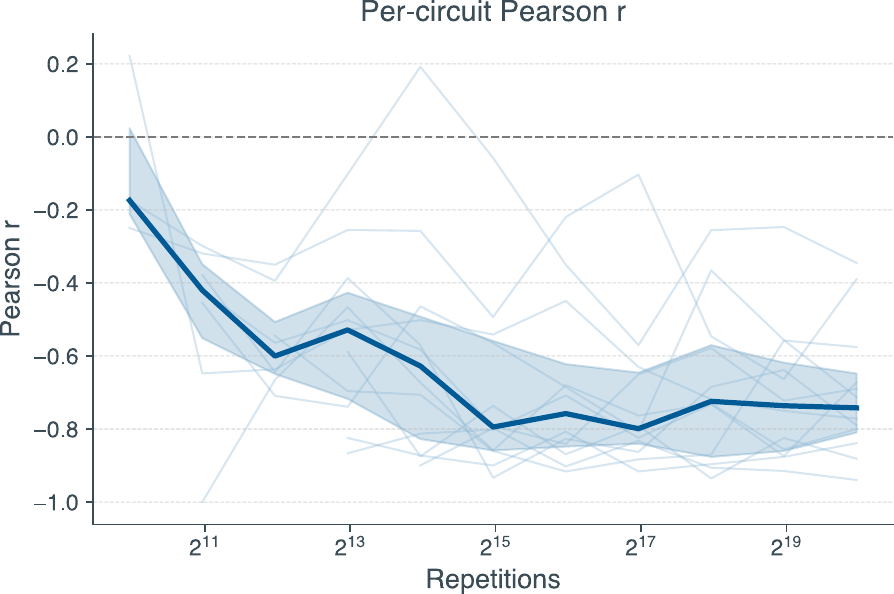}
        \caption{Repetitions per simulation}\label{fig:lsr_tvd_repetitions}
    \end{subfigure}
    \hfill
    \begin{subfigure}{0.48\linewidth}
        \centering
        \includegraphics[width=\linewidth]{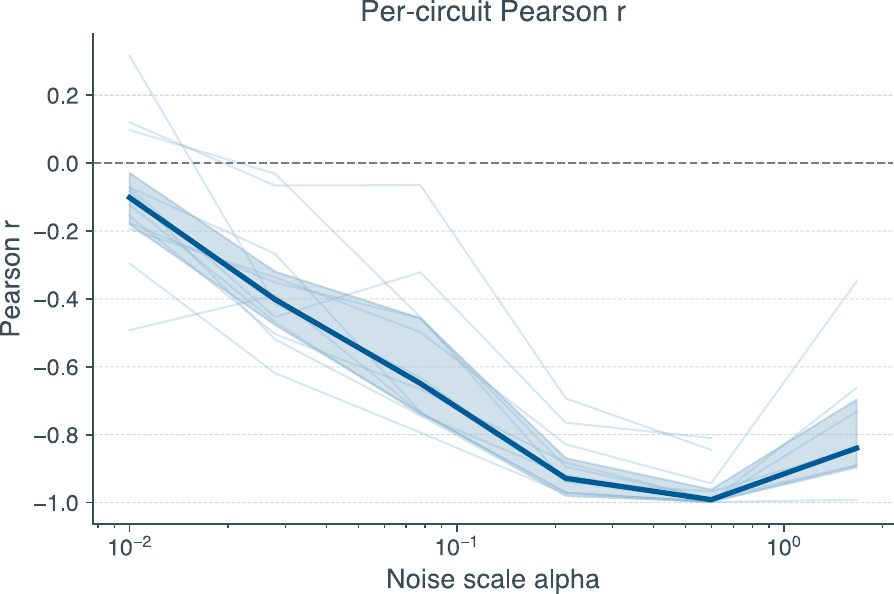}
        \caption{Noise scalar $\alpha$}\label{fig:lsr_tvd_alpha}
    \end{subfigure}
    \caption{\textbf{Logical selection rate and total variation distance are strongly anticorrelated across parameter regimes.} Each subplot sweeps a single parameter over 20 random permutations of 12 random benchmark circuits at $10^5$ shots per permutation. The anticorrelation strengthens with qubit count, tightens with repetitions, and is most pronounced at moderate noise levels.}\label{fig:lsr_tvd_plots}
\end{figure*}

Figure~\ref{fig:lsr_tvd_plots} examines the relationship between LSR and TVD over four parameter sweeps on $100$ random permutations of the \texttt{qec\_en\_n5} benchmark circuit. In each sweep, we observe a strong anti-correlation: permutations that retain more shots after post-selection consistently produce output distributions closer to ideal. 

This coupling arises from the exponential stringency of post-selection on the $[[4,2,2]]$ code. For each Iceberg module, a shot survives only if all parity bits are zero and the 4-bit data block lands on one of $8$ valid codewords out of 16 possible patterns. If outcomes are approximately random, the acceptance probability per module becomes roughly $\frac{1}{4} \cdot \frac{1}{2} = \frac{1}{8}$. Across $m$ modules, this compounds to approximately $\left(\frac{1}{8}\right)^m$, making it exponentially unlikely for a corrupted module to produce a valid-looking but incorrect codeword. At small qubit counts, there is a great enough chance that errors sneak through post-selection as incorrect, but acceptable codewords which degrades TVD even when LSR is moderate. As the number of modules grows, this leakage probability vanishes, and high LSR increasingly implies that the surviving distribution is clean. Therefore LSR and TVD are both downstream effects of the same cause: the total error burden of the circuit. This coupling can be seen in Figure~\ref{fig:lsr_tvd_qubits} with post-selection acting as an exponentially strict filter.

Figure~\ref{fig:lsr_tvd_plots} confirms that this coupling is consistent across simulation parameters. The anti-correlation strengthens towards $-1$ as qubit count increases (Figure~\ref{fig:lsr_tvd_qubits}) and remains mostly constant as qubits remain steady but code count changes (Figure~\ref{fig:lsr_tvd_codes}). This behavior is consistent with the exponential compounding of per-module leakage and repeated distribution reduction on post-selected bit strings that contain untracked qubit readouts. Increasing repetitions (Figure~\ref{fig:lsr_tvd_repetitions}) tightens the observed correlation until it plateaus once sampling noise is sufficiently reduced. Lastly, covariance is strongest when noise is moderate (Figure~\ref{fig:lsr_tvd_alpha}). If noise is too low and post-selection rarely encounters errors which makes TVD an analysis of purely how reflective the sampled distribution is. If noise is too high, then there will be no shots to build a meaningful distribution.

\end{document}